\documentclass[letterpaper,12pt]{article}
\pdfoutput=1

\usepackage{jheppub,framed}
\usepackage{tikz}
\usetikzlibrary{arrows,shapes,decorations.pathmorphing,calc,intersections}
\usepackage{mathtools}
\usepackage{subcaption}

\DeclareCaptionSubType{figure}

\DeclareMathOperator{\Gr}{Gr}
\DeclareMathOperator{\Li}{Li}
\DeclareMathOperator{\conf}{Conf}

\DeclareMathOperator{\spa}{span}%\span is already defined
\DeclareMathOperator{\sgn}{sgn}

\newcommand*\abs[1]{\lvert#1\rvert}
\newcommand*\st[1]{\{#1\}}

\def\braket#1{\mathinner{\langle{#1}\rangle}}
\def\ket#1{\mathinner{\langle{#1}\rangle}}

\title{Hedgehog Bases for $A_n$ Cluster Polylogarithms and An Application
to Six-Point Amplitudes}

\author{Daniel E.~Parker,}
\author{Adam Scherlis,}
\author{Marcus Spradlin}
\author{and Anastasia Volovich}

\affiliation{Department of Physics, Brown University, Providence RI 02912, USA}

\abstract{
Multi-loop scattering amplitudes in
$\mathcal{N}=4$ Yang-Mills theory possess cluster algebra structure.
In order to develop a computational framework which
exploits this connection, we show
how to construct bases of Goncharov polylogarithm functions, at any
weight, whose symbol alphabet consists of cluster coordinates on the
$A_n$ cluster algebra.
Using such a basis we present a new expression for the
2-loop 6-particle NMHV amplitude which makes some of its cluster
structure manifest.
}

\definecolor{red2}{rgb}{1,0.25,0.25}
\definecolor{pur}{rgb}{0.5,0,0.5}
\definecolor{blue2}{rgb}{0.25,0.25,1}
\tikzset{
	vertex/.style={draw,shape=circle,fill=black,minimum size=0pt,inner sep=0pt},
	vertex2/.style={draw,shape=circle,fill=black,minimum size=4pt,inner sep=0pt},
	polygonEdge/.style={draw, very thick,line cap = round},
	blueNode/.style={circle,fill=blue!20,draw},
	factNode/.style={draw,shape=circle,fill=black,inner sep=2pt},
	grayOut/.style={fill=gray!20}
}

\begin{document}
\maketitle

%\linenumbers

\section{Introduction}

In a series of recent papers
following~\cite{Golden:2013xva}
it has been realized that (all known)
multi-loop
$n$-particle
scattering amplitudes of planar $\mathcal{N} = 4$
super-Yang-Mills (SYM) theory possess
special properties that are intimately connected to mathematical structures
known as cluster algebras.  The most basic aspect of this connection
is that amplitudes
are linear combinations of generalized polylogarithm functions
whose symbol arguments are cluster coordinates on the $\Gr(4,n)$
Grassmannian cluster algebra\footnote{This aspect is the focus of
our paper, but other connections have been observed for particular
amplitudes.  For example, there is a tight connection between
the cobracket of motivic 2-loop MHV amplitudes and the Poisson structure
on the underlying cluster
algebra, which has been explored
in~\cite{Golden:2013xva,Golden:2014xqa,Golden:2014pua}.}.

This connection between scattering amplitudes and
cluster algebras is undoubtedly related to
a similar cluster structure that has been observed at the level
of integrands in~\cite{ArkaniHamed:2012nw}, though the precise
connection has yet to be made.
Nevertheless,
the observed cluster structure of integrated
amplitudes has already helped to facilitate the computation of
new expressions for various quantities associated to amplitudes (see for
example~\cite{Golden:2013lha,Golden:2014xqa,Golden:2014xqf,Golden:2014pua,Drummond:2014ffa}).
In parallel, work by Dixon, Drummond, and collaborators has resulted in
spectacular progress
in determining 6-particle amplitudes
via a bootstrap approach (see~\cite{Dixon:2011pw,Dixon:2011nj,Dixon:2013eka,Dixon:2014voa,Dixon:2014iba}, or the review~\cite{Dixon:2014xca}) utilizing input from the OPE of null Wilson loops (see for example~\cite{Basso:2013vsa,Basso:2013aha,Basso:2014koa,Basso:2014nra,Belitsky:2014sla,Belitsky:2014lta}).

Typically, results in SYM theory take the form of colossal linear
combinations of generalized polylogarithm functions.
These special functions satisfy a huge number of functional identities: shuffle identities, stuffle identities, the Abel identity, the trilogarithm identity
of~\cite{Golden:2013xva}, and many others.
These make generalized polylogarithms notoriously difficult to work with. Moreover, with so many identities, there are a multitude of possible ways to write the same formula.  In general there is no ``best'' way to write
a given expression, nor is it even clear how one ought to define ``best'' --- perhaps the shortest expression, or one where certain physical or mathematical properties are manifest.

Large progress towards finding canonical bases for generalized polylogarithms has been made by Brown in~\cite{Brown} (see also~\cite{Bogner:2014mha} for some applications)
and employed by Dixon et.~al.~in their 6-particle bootstrap program.
In this paper we demonstrate a natural way to ``clusterize'' Brown's
basis of polylogarithm functions.
Namely, we show how to generate, at any weight, a basis of generalized
polylogarithm functions whose symbols are manifestly expressible in terms of
cluster coordinates on the
$A_n$ cluster algebra.
We call these ``hedgehog'' bases because they are
naturally associated to certain spiny structures in the $A_n$ exchange graph.
Hedgehog bases provide an almost canonical way to write expressions for
6-particle MHV and NMHV amplitudes, presumably
at any loop order.  Compared to using
other bases that have been considered in the literature, hedgehog
bases have the theoretically-pleasing advantage of making some of the cluster
structure of such amplitudes manifest, as well as the practical benefit
of allowing notably shorter expressions.
The latter feature echoes a common theme in the amplitudes program:
identifying underlying mathematical structure and improving
computational efficiency go hand in hand.

Section~2 briefly reviews the necessary mathematical technology of polylogarithms, cluster algebras, and scattering amplitudes. Section~3 introduces the idea of a ``hedgehog'' for a cluster algebra, and sketches the rigorous proof (with details relegated to an appendix) that they can be fashioned into a basis for polylogarithms.
Section~4 presents, as an application of this technology, a construction of
a hedgehog-basis representation for the the 2-loop 6-particle NMHV amplitude\footnote{Our result is included as an ancillary file with the arXiv submission.}.

\section{Review}
Polylogarithms and cluster algebras are each subjects unto themselves.
Thus this section is not an all-encompassing review, but rather a brief
reminder of some of the mathematical technology needed for the rest of the
paper, together with citations where the curious reader may find additional
details.  We also review the relevant aspects of the connection between
Grassmannian cluster algebras and scattering amplitudes in SYM theory.

\subsection{Generalized Polylogarithms}
Polylogarithms are a broad class of special functions that generalize the logarithm. More details on the material in this section may be found in the recent review~\cite{Duhr:2014woa}.

Recall that the ordinary logarithm can be written as $\log z = \int_0^{z} \frac{dt}{t}$. Generalizing this to an iterated integral of the type first studied systematically by Chen~\cite{Chen} gives the \textbf{weight-$k$ Goncharov polylogarithm}~\cite{Gfunctions}:
\begin{equation}
	G(a_1, \dots, a_k; z) = \int_0^z \frac{dt}{t-a_1} G(a_2, \dots, a_k; t), \qquad G(z) = 1,
	\label{eq:Gdef}
\end{equation}
with the special case
\begin{equation}
\label{eq:specialcase}
	G(\underbrace{0,\ldots,0}_k; z) = \frac{1}{k!} \log^k z.
\end{equation}
In general, $a_1, \dots, a_k$ are valued in the complex numbers with $z \in \mathbb{C}\setminus \st{a_1, \dots, a_k}$, and one should specify a contour of integration. We will see that for scattering amplitudes in a certain domain these variables are all real-valued, and there is a natural ordering which allows one to take the ``naive'' contour straight along the real axis. A large class of $L$-loop amplitudes in SYM theory, including at least all MHV and NMHV amplitudes, are expected to be expressible as linear combinations of weight-$2L$ polylogarithms. The classical polylogarithms $\Li_k(z) = - G(\underbrace{0,\dots, 0}_{k-1}, 1; z)$ form a strict subset of the Goncharov polylogarithms.

As mentioned in the introduction, the bane of working with polylogarithms is the numerous functional identities they obey.
Perhaps the most important of these is the \textbf{shuffle identity}. The product of two polylogarithms can be written as
\begin{equation}
	G(a_1, \dots, a_n; z) G(a_{n+1}, \dots, a_{n+m}; z) = \sum_{\sigma\in S_{n,m}} G(a_{\sigma(1)}, \dots, a_{\sigma(n+m)}; z)
	\label{eq:shuffle_identity}
\end{equation}
where $S_{n,m}$ is the set of $(n,m)$-shuffles, i.e. permutations $\sigma$ of length $n+m$ such that
\begin{equation}
	\sigma^{-1}(1) < \sigma^{-1}(2) < \cdots < \sigma^{-1}(n) \text{ and } \sigma^{-1}(n+1) < \sigma^{-1}(n+2) < \cdots < \sigma^{-1}(n+m).
	\label{eq:shuffle_permutation}
\end{equation}
The name comes from riffle shuffling a deck of cards; shuffling two stacks of cards together interweaves them while leaving each stack in the same order.

Each polylogarithm has an associated object called its \textbf{symbol} (see for example~\cite{Goncharovsimple,Goncharov:2010jf}, and the review~\cite{Duhr:2011zq}). 
The symbol is a useful tool for converting the functional identities of polylogarithms into linear algebra, obviating many thorny problems. The symbol of a Goncharov polylogarithm admits a nice graphical interpretation as a sum over plane trivalent trees~\cite{GoncharovHopf}, and is given explicitly by the recursive formula
\begin{align}
\begin{split}
S(G(a_k,\ldots,a_1;a_{k+1})) &= \sum_{i=1}^k
S(G(a_{k},\ldots,\widehat{a}_i,\ldots,a_1;a_{k+1})) \otimes (a_i{-}a_{i+1})\\
&\qquad - S(G(a_{k},\ldots,\widehat{a}_i,\ldots,a_1;a_{k+1})) \otimes (a_i{-}a_{i-1}).
\end{split}
\label{eq:Gsymbol}
\end{align}
Here $\widehat{a}_i$ denotes that the argument is omitted, and
it is also understood that any term with 0 as a symbol entry (which can happen if some adjacent $a$'s are equal) should simply be omitted.
Symbols behave as if there were implicit ``$d \log$'s'' in front of each term: just as $d \log 1 = 0$ and $d \log \phi_1\phi_2 = d \log \phi_1 + d \log \phi_2$, symbols obey $( \cdots \otimes 1 \otimes \cdots ) = 0$ and
\begin{equation}
\label{eq:symbolexpand}
( \alpha \otimes \phi_1 \phi_2 \otimes \beta )
= ( \alpha \otimes \phi_1 \otimes \beta )
+ ( \alpha \otimes \phi_2 \otimes \beta ).
\end{equation}
The collection of $\phi_i$ which appear in the symbol of a given function is called its \textbf{symbol alphabet}.

\subsection{Cluster Algebras}
\label{sec:cluster_algebras_review}

Cluster algebras are a relatively new area of mathematics, introduced in 2002 by Fomin and Zelevinsky in~\cite{1021.16017,1054.17024}. This section quickly reviews some salient facts about cluster algebras; the reader may consult~\cite{GSV,Williams} for additional mathematical background and~\cite{Golden:2013xva} for the amplitude perspective.
A cluster algebra starts with a \textbf{seed} --- a quiver where each vertex is labeled with a \textbf{cluster variable} (also called a \textbf{cluster coordinate})\footnote{To be clear, throughout this paper the term ``cluster variable'' refers to the $\mathcal{X}$-coordinates of Fock and Goncharov~\cite{FG03b}, not to cluster $\mathcal{A}$-coordinates.}. See figure~\ref{fig:A3_initial_quiver} for an example of a seed.

\begin{figure}[h]
	\center
	\begin{subfigure}[b]{0.5\textwidth}\center
\begin{tikzpicture}[->,>=stealth',shorten >=1pt,auto,node distance=2cm,
  very thick]
 		\node[blueNode] (1) at (0,0) {$x_1$};
		\node[blueNode] (2) [right of=1] {$x_2$};
		\node[blueNode] (3) [right of=2] {$x_3$};
		\draw[->] (1) -- (2);
		\draw[->] (2) -- (3);
\end{tikzpicture}
\caption{The initial seed for the $A_3$ algebra.}
\label{fig:A3_initial_quiver}
\end{subfigure}%leave this comment here to kill the linebreak
\begin{subfigure}[b]{0.5\textwidth}\center
\begin{tikzpicture}[->,>=stealth',shorten >=1pt,auto,node distance=2cm,
  very thick]
  \node[blueNode] (1) at (0,0) {$x_1'$};
		\node[blueNode] (2) [right of=1] {$\tfrac{1}{x_2}$};
		\node[blueNode] (3) [right of=2] {$x_3'$};
		\draw[->] (2) -- (1);
		\draw[->] (3) -- (2);
		\draw[->] (1) to[bend left=45] (3);
\end{tikzpicture}
\caption{The result of mutating on the vertex $x_2$.}
\label{fig:A3_mut_quiver}
\end{subfigure}
\caption{}
\end{figure}

A cluster algebra is generated from an initial seed through an iterative process. An operation called \textbf{mutation on a vertex} generates a new seed and new cluster variables, according to the formula given in eq.~\eqref{eq:mutationrule} (or see~\cite{FG03b}). For example, mutating on the middle vertex on figure~\ref{fig:A3_initial_quiver} gives figure~\ref{fig:A3_mut_quiver} with the new cluster variables $x_1' = x_1(1+x_2)$ and $x_3' = \frac{x_2 x_3}{1+x_2}$. Mutation is an involution, so applying the same mutation twice does nothing. The \textbf{cluster algebra} is the algebra generated by the set of all cluster variables which arise from repeatedly mutating the initial seed. Under certain conditions on the initial quiver, all possible repeated mutations will yield only finitely many seeds. Such cluster algebras are said to be of \textbf{finite type}.

A natural domain for a cluster algebra is the
\textbf{positive domain}, where all cluster variables take positive real
values.  This property is preserved under mutation: if all variables in
a given seed are positive-valued, then all possible cluster variables on
the same algebra are also positive-valued.

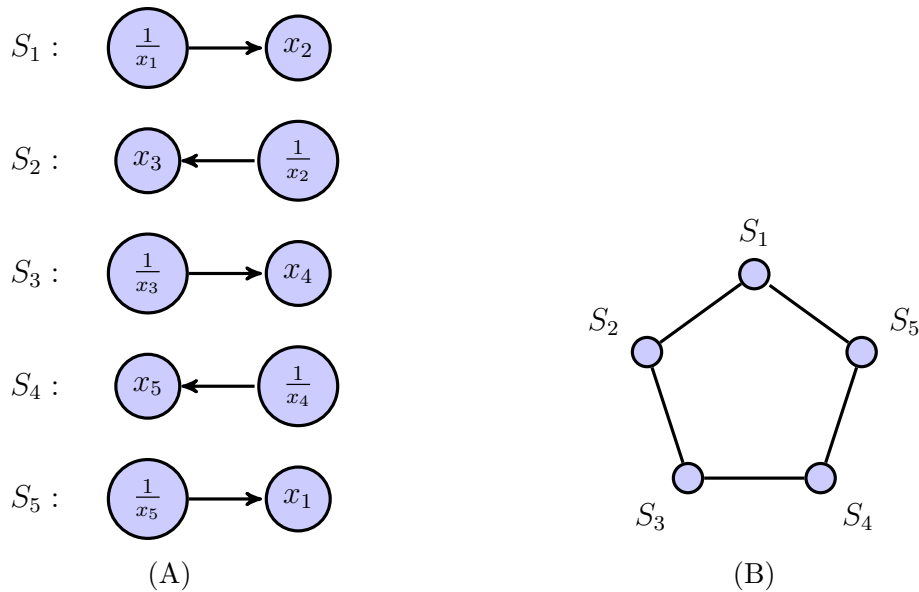
\begin{figure}[t]
	\center
	\begin{subfigure}[b]{0.5\textwidth}\centering
	\begin{tikzpicture}[->,>=stealth',shorten >=1pt,auto,node distance=2cm,
  very thick]
  		\foreach \i in {1,...,5}{\node[] at (-1.5, 7.5-1.5*\i) {$S_\i:$};}

		\node[blueNode] (l1) at (0,6) {$\frac{1}{x_1}$};
		\node[blueNode] (l2) at (0,4.5) {${x_3}$};
		\node[blueNode] (l3) at (0,3) {$\frac1{x_3}$}; 
		\node[blueNode] (l4) at (0,1.5) {$x_5$};
		\node[blueNode] (l5) at (0,0) {$\frac{1}{x_5}$};

		\node[blueNode] (r1) at (2,6) {$x_2$};
		\node[blueNode] (r2) at (2,4.5) {$\frac{1}{x_2}$};
		\node[blueNode] (r3) at (2,3) {$x_4$}; 
		\node[blueNode] (r4) at (2,1.5) {$\frac{1}{x_4}$};
		\node[blueNode] (r5) at (2,0) {$x_1$};

		\draw[->] (l1) -- (r1);
		\draw[<-] (l2) -- (r2);
		\draw[->] (l3) -- (r3);
		\draw[<-] (l4) -- (r4);
		\draw[->] (l5) -- (r5);
	\end{tikzpicture}
	\caption{}
	\label{fig:A2_seeds}
	\end{subfigure}%
	\begin{subfigure}[b]{0.5\textwidth}\centering
	\begin{tikzpicture}[->,>=stealth',shorten >=1pt,auto,node distance=2cm,
  very thick]
		\foreach \i in {1,...,5}{\node[blueNode,label={18+72*\i:$S_\i$}] (p\i) at (18+72*\i:1.5) {};}
		\draw[-] (p1) -- (p2) -- (p3) -- (p4)  -- (p5) -- (p1);
	\end{tikzpicture}
	\caption{}
	\label{fig:A2_exchange_graph}
	\end{subfigure}
	\caption{(A) The five seeds for the $A_2$ algebra, with each node labeled
by its associated cluster variable.  (B) The exchange graph, showing how to move from
one seed to another by mutation.}
\end{figure}

The structure of the cluster algebra as a whole can be displayed as an \textbf{exchange graph}, where each vertex represents a seed, and \textit{undirected} edges are drawn between seeds linked by a single mutation. (See figure~\ref{fig:A2_exchange_graph}.) Because applying a mutation will invert a single cluster variable $x_i \mapsto \frac{1}{x_i}$, a \textit{directed} edge of the exchange graph can be associated to a unique cluster variable $x_i$. The same edge with the opposite direction corresponds to $1/x_i$.

Later in this paper we will focus on the $A_n$ family of cluster algebras, which start with the initial seed
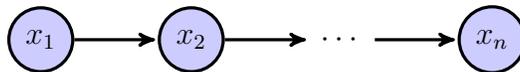
\begin{center}
\begin{tikzpicture}[->,>=stealth',shorten >=1pt,auto,node distance=2cm,
  very thick]
 		\node[blueNode] (1) at (0,0) {$x_1$};
		\node[blueNode] (2) at (2,0) {$x_2$};
		\node[] (dots) at (4,0) {$\cdots$};
		\node[blueNode] (3) at (6,0) {$x_n$};
		\draw[->] (1) -- (2);
		\draw[->] (2) -- (dots);
		\draw[->] (dots) -- (3);
\end{tikzpicture}
\end{center}
for $n \ge 1$. For the special case of $A_n$ algebras, which are all
of finite type,
there is a convenient alternative to representing clusters with quivers.  This construction is reviewed
in appendix~\ref{app:triangulations}.

The $A_2$ algebra, for example, has exactly 5 distinct seeds and 10 cluster variables\footnote{In this paper, we treat $x$ and $1/x$ as two separate cluster variables, but they are sometimes conflated in the literature when it is useful to do so.} given by
\begin{equation}
\label{eq:A2coordinates}
x_1, \quad
x_2, \quad
x_3 = \frac{1 + x_2}{x_1}, \quad
x_4 = \frac{1 + x_1 + x_2}{x_1 x_2}, \quad
x_5 = \frac{1 + x_1}{x_2}
\end{equation}
and their reciprocals.
These variables obey the recursive formula
\begin{equation}
	x_{i+1} = \frac{1+x_i}{x_{i-1}}.
	\label{eq:A2_recursion}
\end{equation}

This paper makes extensive use of the $A_3$ cluster algebra, which starts with the initial seed in figure~\ref{fig:A3_initial_quiver}. It has 14 seeds and 30 cluster variables. We take the opportunity to enumerate in eq.~\eqref{eq:A3coords} 15 of these cluster variables (the other 15 are their reciprocals) in four ways:  (1) in terms of the names $v_i$, $x^\pm_i$ and $e_i$ that these variables have been given in previous work (see in particular~\cite{Golden:2013xva,Golden:2014xqa}), (2) as rational functions of the variables $x_1, x_2, x_3$ in the initial seed, (3) in terms of the $\{u, v, w, y_u, y_v, y_w\}$ variables used extensively by Dixon et.~al.\ in their study of 6-particle scattering amplitudes in SYM theory, (4) and in terms of Pl\"ucker coordinates on $\Gr(4,6)$ (the connection to Pl\"ucker coordinates is explained in the
following subsection)\footnote{The $A_3$ algebra has several seeds
of the form shown in figure~\ref{fig:A3_initial_quiver}.
We caution that
the definition of $x_1$, $x_2$ and $x_3$ in eq.~(\ref{eq:A3coords})
reflects a convention chosen in~\cite{Golden:2014xqa} which differs from
the particular $x_1$, $x_2$ and $x_3$ assigned to the initial seed
in~\cite{GSV}.}

\begin{equation}
	\begin{matrix*}[l]
\label{eq:A3coords}
v_1 &= \tfrac{(1 + x_2)(1 + x_3 + x_2 x_3 + x_1 x_2 x_3)}{x_1 x_2} &= \dfrac{1-v}{v} &= \dfrac{\ket{1246}\ket{1345}}{\ket{1234}\ket{1456}} \\
v_2 &= \dfrac{1 + x_3}{x_2 x_3} &= \dfrac{1-w}{w} &= \dfrac{\ket{1235}\ket{2456}}{\ket{1256}\ket{2345}} \\
v_3 &= (1 + x_1)x_2 &= \dfrac{1-u}{u} &= \dfrac{\ket{1356}\ket{2346}}{\ket{1236}\ket{3456}} \\
x_1^+ &= \dfrac{1}{x_3} &= \sqrt{\dfrac{v y_u y_v y_w}{u w}} &= \dfrac{\ket{1456}\ket{2356}}{\ket{1256}\ket{3456}} \\
x_2^+ &= \dfrac{1 + x_2 + x_1 x_2}{x_1} &= \sqrt{\dfrac{w y_u y_v y_w}{u v}} &= \dfrac{\ket{1346}\ket{2345}}{\ket{1234}\ket{3456}} \\
x_3^+ &= \dfrac{1 + x_3 + x_2 x_3}{x_1 x_2 x_3} &= \sqrt{\dfrac{u y_u y_v y_w}{v w}} &= \dfrac{\ket{1236}\ket{1245}}{\ket{1234}\ket{1256}} \\
x_1^- &= x_1 &= \sqrt{\dfrac{v}{u w y_u y_v y_w}} &= \dfrac{\ket{1234}\ket{2356}}{\ket{1236}\ket{2345}} \\
x_2^- &= (1 + x_2 + x_1 x_2) x_3 &= \sqrt{\dfrac{w}{u v y_u y_v y_w}} &= \dfrac{\ket{1256}\ket{1346}}{\ket{1236}\ket{1456}} \\
x_3^- &= \dfrac{1 + x_3 + x_2 x_3}{x_2} &= \sqrt{\dfrac{u}{v w y_u y_v y_w}} &= \dfrac{\ket{1245}\ket{3456}}{\ket{1456}\ket{2345}} \\
e_1 &= \dfrac{1 + x_3 + x_2 x_3 + x_1 x_2 x_3}{(1 + x_1) x_2} &= \sqrt{\dfrac{(1-v) u}{v (1-u) y_u y_v}} &= \dfrac{\ket{1246}\ket{3456}}{\ket{1456}\ket{2346}} \\
e_2 &= \dfrac{1}{(1 + x_2) x_3} &= \sqrt{\dfrac{v (1-w) y_v y_w}{(1-v) w}} &= \dfrac{\ket{1235}\ket{1456}}{\ket{1256}\ket{1345}} \\
e_3 &= \dfrac{(1 + x_1) x_2 x_3}{1 + x_3} &= \sqrt{\dfrac{w(1-u)}{(1-w) u y_u y_w}} &= \dfrac{\ket{1256}\ket{2346}}{\ket{1236}\ket{2456}} \\
e_4 &= \dfrac{1 + x_2}{x_1 x_2} &= \sqrt{\dfrac{(1-v) u y_u y_v}{v(1-u)}} &= \dfrac{\ket{1236}\ket{1345}}{\ket{1234}\ket{1356}} \\
e_5 &= \dfrac{x_1 (1 + x_3)}{1 + x_3 + x_2 x_3 + x_1 x_2 x_3} &= \sqrt{\dfrac{v(1-w)}{(1-v) w y_v y_w}} &= \dfrac{\ket{1234}\ket{2456}}{\ket{1246}\ket{2345}} \\
e_6 &= x_2 &= \sqrt{\dfrac{w (1-u) y_u y_w}{(1-w) u}} &= \dfrac{\ket{1356}\ket{2345}}{\ket{1235}\ket{3456}}
\end{matrix*}
\end{equation}

\subsection{Scattering Amplitudes and Grassmannian Cluster Algebras}

The connection between scattering amplitudes in SYM theory
and cluster algebras was first made in~\cite{Golden:2013xva}
and further explored in~\cite{Torres:2013vba,Golden:2014xqa,Paulos:2014dja,Golden:2014pua}.
The basic fact that allows for such a connection is that
the kinematic domain for $n$-particle scattering in SYM theory,
called $\conf_n(\mathbb{P}^3)$, has,
according to~\cite{FG03b}, the structure of a cluster Poisson variety
associated to the $\Gr(4,n)$ Grassmannian cluster algebra.
This fact is special to SYM theory in four dimensions because it relies
on the dual conformal symmetry of the theory, discovered
in~\cite{Drummond:2008vq,Drummond:2007au,Bern:2007ct,Alday:2007hr,Bern:2006ew,Drummond:2006rz}.

An ordered scattering amplitude of $n$ massless particles is a function
of $n$ null vectors in Minkowski space that sum up to zero
due to energy-momentum conservation.  Using the momentum twistor
variables of Hodges~\cite{Hodges:2009hk}, the space of such configurations
can be realized as $n$ ordered points in $\mathbb{P}^3$, or concretely
as a $4 \times n$ matrix $[Z_1 Z_2 \cdots Z_n]$ where each column
$Z_i$ is a four-component homogeneous coordinate on $\mathbb{P}^3$.
In this presentation, dual conformal symmetry, which must leave all
amplitudes invariant, acts as left-multiplication by $SL(4,\mathbb{C})$.
Passing to the quotient space we get a birational isomorphism
(which means a bijection for generic points)
\begin{equation}
\Gr(4, n)/(\mathbb{C}^*)^{n-1}
\stackrel{\sim}{\longrightarrow} \conf_n(\mathbb{P}^{3}).
\end{equation}

Thus, scattering amplitudes can be (essentially) regarded as complex-valued
functions on Grassmannians, making it natural to use
the $SL(4,\mathbb{C})$ invariant Pl\"ucker coordinates
$\langle ijk\ell \rangle = \det[Z_i Z_j Z_k Z_\ell]$, which are well-defined
complex-valued functions on $\Gr(4,n)$.  However, since the $Z$'s are
homogeneous coordinates, it is necessary to use
ratios of Pl\"ucker coordinates (or, more generally, ratios of
homogeneous polynomials of Pl\"ucker coordinates), with the same $Z$'s
appearing in the numerator and denominator, such as
\begin{equation}
        \frac{\braket{5713}\braket{5624}}{\braket{4512}\braket{3567}},
        \label{eq:example_cross_ratio}
\end{equation}
to get well-defined coordinates on $\mathrm{Gr}(4,n)/(\mathbb{C^*})^{n-1}$.
Scattering amplitudes in SYM theory are
naturally written as functions of such cross-ratios.

This is where cluster algebras enter:
the Pl\"ucker coordinates of any Grassmannian form
a cluster algebra~\cite{Scott}, and the quotient
$\conf_n(\mathbb{P}^3)$ has the structure of a cluster
Poisson variety~\cite{FG03b},
with cluster coordinates given by certain very special cross-ratios
of the abovementioned type\footnote{The positive domain for the $\Gr(4,n)$ algebra is defined by $\ket{ijk\ell} > 0~\forall 1 \le i<j<k<\ell\le n$.}.
The physics interest in such cluster algebras stems from the fact that
all known multi-loop amplitudes that have been explicitly computed
to date in SYM theory
(including~\cite{DelDuca:2009au,DelDuca:2010zg,Goncharov:2010jf,CaronHuot:2011ky,Dixon:2011nj,CaronHuot:2011kk,Golden:2013xva,Dixon:2013eka,Dixon:2014voa,Dixon:2014iba})
are generalized polylogarithms whose symbol alphabets
are subsets of cluster coordinates on this $\Gr(4,n)$
cluster algebra\footnote{A number of results in two-dimensional kinematics including~\cite{DelDuca:2010zp,Heslop:2010kq,Heslop:2011hv,Goddard:2012cx,Caron-Huot:2013vda} provide partial evidence in support this assertion, though the full $\Gr(4,n)$ structure necessarily collapses in two-dimensional kinematics. This has been studied in~\cite{Torres:2013vba}.}.
The $A_n$ family of cluster algebras reviewed in section~\ref{sec:cluster_algebras_review}
corresponds to the Grassmannian $\Gr(2,n+3)$, which overlaps with the sequence
of algebras relevant to scattering amplitudes in the case
$A_3 \cong \Gr(2,6) \cong \Gr(4,6)$ relevant to 6-particle amplitudes.

\subsection{The Cluster Bootstrap}

The main problem we address
in this paper
is simple to state:  given a cluster algebra $\mathcal{A}$,
with a set of cluster coordinates $\mathcal{X}_\mathcal{A}$,
we would like to
write down a basis
for weight-$k$ polylogarithm functions whose symbols
may be written in the alphabet $\mathcal{X}_\mathcal{A}$.
We call such functions
``cluster polylogarithm functions'' or simply
{\bf cluster functions}\footnote{Functions of this type were called ``cluster $\mathcal{A}$-functions''
in~\cite{Golden:2014xqa} to distinguish them from a smaller set of
functions with more special properties called
``cluster $\mathcal{X}$-functions'', but we do not
explore these additional properties here.}
on $\mathcal{A}$.

To be explicit, let us note
that $A_2$ cluster functions, for example, are those which can be written
in the symbol alphabet consisting of the five $x_i$ shown in
eq.~\eqref{eq:A2coordinates}.  Thanks to eq.~\eqref{eq:symbolexpand},
we can equivalently consider the $A_2$ symbol alphabet to be the set
\begin{equation}
\{ x_1, x_2, 1 + x_1, 1 + x_2, 1 + x_1 + x_2 \}
\end{equation}
since each of the five $x_i$ may be (uniquely) expressed as products of powers
of elements of this set.  For $A_3$ only 9 of the 30 cluster variables
are multiplicatively independent, and it is evident from
eq.~\eqref{eq:A3coords}
that the $A_3$ symbol alphabet may be taken as the set
\begin{equation}
\{x_1, x_2, x_3, 1 + x_1, 1 + x_2, 1 + x_3,
1 + x_2 + x_1 x_2, 1 + x_3 + x_2 x_3,
1 + x_3 + x_2 x_3 + x_1 x_2 x_3\}.
\label{eq:A3alphabet}
\end{equation}

Closely related symbol alphabets have appeared elsewhere,
notably in Brown's work on polylogarithm functions on the moduli
space $\mathfrak{M}_{0,m}$ of $m$ marked points on the Riemann
sphere~\cite{Brown}.
For example, for the case $m=6$, Brown's polylogarithms are based on
the symbol alphabet
\begin{equation}
\label{eq:brownlettersone}
\{ c_1, c_2, c_3, 1 - c_1, 1 - c_2, 1 - c_3,
1 - c_1 c_2,
1 - c_2 c_3,
1 - c_1 c_2 c_3 \}
\end{equation}
in cubical coordinates\footnote{These cubical coordinates
were called $x_i$
in~\cite{Brown}, but we use $c_i$ in eq.~(\ref{eq:brownlettersone}) to
distinguish them from our $x_i$ cluster coordinates.} or
\begin{equation}
\label{eq:brownletterstwo}
\{ t_1, t_2, t_3, 1 - t_1, 1 - t_2, 1 - t_3,
t_3 - t_1,
t_3 - t_1,
t_3 - t_2 \}
\end{equation}
in simplicial coordinates.
Neither alphabet is multiplicatively equivalent to
eq.~(\ref{eq:A3alphabet}), but their relation will be uncovered in
the following section.  
In fact, one way to express the central result of our paper is to
say that we demonstrate
how to construct explicit changes of variables
between those of~\cite{Brown}
on $\mathfrak{M}_{0,n+3}$ and the $A_n$
cluster $\mathcal{X}$-coordinates, for any $n$,
which render the
corresponding symbol alphabets multiplicatively
equivalent.

Let us conclude our review by briefly recalling that
for finite symbol alphabets this problem
admits a conceptually straightforward, if
computationally intensive, brute force solution.
If the symbol alphabet for $\mathcal{A}$ has $s$ multiplicatively independent
letters $\{\phi_1,\ldots,\phi_s\}$,
then the symbol of any weight-$k$ cluster function may
be expressed as a unique vector (with rational components)
in the $s^k$ dimensional vector space $\mathcal{V}_k$ spanned
by basis elements $\phi_{i_1} \otimes \cdots \otimes \phi_{i_k}$.
Going the other way around,
any vector in $\mathcal{V}_k$ which satisfies a set of
linear
integrability conditions (see for example~\cite{Goncharovsimple})
corresponds to (the symbol of) some cluster function.
Therefore, the problem of finding a basis for the (symbols of)
weight-$k$ cluster functions
on $\mathcal{A}$ is the same as that finding a basis for the nullspace of
a certain linear operator on $\mathcal{V}_k$.

The efficiency of this approach can be considerably enhanced by
recycling lower-weight information at higher weight, and by exploiting
the Hopf algebra structure of polylogarithms (discovered in~\cite{GoncharovHopf}, and nicely reviewed for a physics audience in~\cite{Duhr:2012fh})\footnote{The Hopf algebra structure makes SYM theory an ideal setting in which to study \emph{motivic amplitudes}, as proposed a decade ago in~\cite{GoncharovHopf} (see in particular section~7).}.
Collectively these ``bootstrap'' techniques have been implemented
systematically
by Dixon and collaborators for the 6-particle case (associated
to the $\Gr(4,6) \cong A_3$ cluster algebra)
to great effect
in~\cite{Dixon:2013eka,Dixon:2014voa,Dixon:2014iba}.
A slightly modified ``weight-skipping'' bootstrap based on
a symbol alphabet of $\Gr(4,7)$ cluster coordinates allowed
for the calculation of the symbol of the 3-loop 7-particle MHV
amplitude
in~\cite{Drummond:2014ffa}.

Finally, we note a fact we will use later:
the classical polylogarithm functions $\Li_k$ (and
products thereof) are known to span the space of all polylogarithm
functions of weight $k \le 3$, so it is trivial to write down a (vastly
overcomplete) set of irreducible cluster functions at weights $k=1,2,3$:
\begin{equation}
\label{eq:trivialclusterfunctions}
\log(\mathcal{X}_\mathcal{A}), \quad
\Li_2(-\mathcal{X}_\mathcal{A}), \quad
\Li_3(-\mathcal{X}_\mathcal{A}) \cup
\Li_3(1 + \mathcal{X}_\mathcal{A}).
\end{equation}
The problem we address in this paper is that of finding bases for all
weights, not just
overcomplete sets of cluster functions.

\section{Hedgehog Bases}

We tackle the problem of constructing bases of cluster functions in three steps.
(1)~First we discuss the set of Goncharov polylogarithms whose symbols may be written in the alphabet of cluster coordinates.
(2)~Next, we review the form a generating set should have, based on work of
Brown~\cite{Brown} and Drummond~\cite{JD}.
(3)~Lastly, we define ``hedgehogs'' and prove that they provide bases for the space of $A_n$ cluster functions.

\subsection{Good Arguments for Goncharov Polylogarithms}
\label{sec:good_arguments}

To construct suitable collections of functions there is no need
to reinvent the wheel.
We may
attempt to solve this problem by using a nice set of polylogarithm
functions we already have at our disposal: the Goncharov polylogarithms defined
in eq.~(\ref{eq:Gdef}).
Then it remains only to decide what kinds of variables we should allow
as the arguments $a_1,\ldots,a_k;z$.

Let us write $G_k[Q]$ to denote the set of weight-$k$
Goncharov polylogarithms whose arguments are drawn from some set $Q$:
\begin{equation}
\label{eq:Gkdef}
G_k[Q] = \{ G(q_1,\ldots,q_k; q_0) : q_i \in Q \}.
\end{equation}
It is evident from eq.~(\ref{eq:Gsymbol}) that functions in
$G_k[Q]$ have symbol entries of the form
$q_i$ as well as
$q_i{-}q_j$, for $q_i, q_j \in Q$.
We may try to follow the path of least resistance by considering what happens
when $Q$ is chosen simply
to be some subset of $\mathcal{X}_\mathcal{A}$.
Actually, although this doesn't matter at the level of
symbols, for later convenience it will be better to consider subsets
of $-\mathcal{X}_\mathcal{A}$ since this will help to naturally provide
Goncharov polylogarithms that are manifestly free of branch cuts in the positive domain.
(Henceforth we shall use $x_i \in \mathcal{X}_\mathcal{A}$
to denote cluster coordinates
and $q_i = - x_i$ to denote negative cluster coordinates.)
Unfortunately,
for two generic $q_i, q_j \in -\mathcal{X}_\mathcal{A}$,
there is nothing particularly nice about the quantity $q_i{-}q_j$;
it may not even have definite sign in the positive domain, in which case it 
should never appear in the symbol of a cluster function.

One approach to construct bases of cluster functions would use
special linear combinations of Goncharov polylogarithms for which all
``bad'' letters cancel out at the level of symbols.
Several examples of such functions have been studied in the literature.
For the particular case of $\mathcal{A} = A_3$,
Dixon et.~al.~have constructed Goncharov polylogarithm representations
for bases of ``hexagon functions'' through weight at least~8.
These are cluster functions satisfying an additional important physical constraint
(the first-entry condition), which we do not address here.
The construction of these bases, and several impressive applications to
6-particle scattering
amplitudes in SYM theory, are discussed
in~\cite{Dixon:2011nj,Dixon:2013eka,Dixon:2014voa,Dixon:2014xca,Dixon:2014iba}.
Also, the ``cluster $\mathcal{X}$-functions''
studied in~\cite{Golden:2014xqa,Golden:2014xqf} for
more general algebras can be expressed as suitable linear combinations
of Goncharov polylogarithms with all ``bad'' symbol entries cancelling out.
These functions also play a prominent role in SYM theory: in particular,
it appears from the result
of~\cite{Golden:2014pua}
that all 2-loop MHV amplitudes can be expressed
in terms of classical polylogarithms and the
single non-classical
cluster function $K_{2,2}$ defined in~\cite{Golden:2014xqf}.

In the present paper, we would like to explore
a different approach to cluster functions.
We explore the possibility of
constructing Goncharov polylogarithms at any
weight which are manifestly free of any
``bad'' letters, rather than having to rely
on solving a (potentially computationally-challenging) linear algebra problem
to ensure their cancellation.
In light of the factorization property reviewed in eq.~(\ref{eq:symbolexpand}),
it is evident that
this will be the case if we can choose the set $Q$ so that $q_i{-}q_j$
factors into a product of powers of
cluster coordinates for all $q_i,q_j \in Q$.
To be precise, let us
define the {\bf multiplicative span}
of $\mathcal{X}_\mathcal{A}$ to be the set
\begin{equation}
M_\mathcal{A} = \{ \pm \prod_i x_i^{n_i} : x_i \in \mathcal{X}_\mathcal{A}
\text{ and } n_i \in \mathbb{Z} \}.
\end{equation}
(If $\mathcal{A}$ is an infinite algebra, then
only finitely many of the $n_i$ may be nonzero.)
We say that a set $Q$ {\bf splits over} $\mathcal{X}_\mathcal{A}$
if $q_i{-}q_j \in M_\mathcal{A}$ for all $q_i \ne q_j \in Q$.

Then it is evident that $G_k[Q]$ is a set of cluster
functions on $\mathcal{A}$ whenever $Q \subseteq - \mathcal{X}_\mathcal{A}$
splits over $\mathcal{X}_\mathcal{A}$.
In fact, for any such $Q$ we can get
additional cluster functions ``for free'' by considering
the enlarged set $G_k[\{0,1\} \cup Q]$.
The inclusion of 0 is trivial,
and 1 is allowed because of the property that
$q - 1 = 1 + x\in M_\mathcal{A}$ for all $q \in - \mathcal{X}_\mathcal{A}$.
A proof of this property, which played an important
role in~\cite{Golden:2013xva,Golden:2014xqf,Golden:2014pua},
is presented in appendix~\ref{app:theorem}.

We can conclude that

\begin{center}
\begin{framed}
\noindent
If $Q \subseteq -\mathcal{X}_\mathcal{A}$
splits over $\mathcal{X}_\mathcal{A}$, then
$G_k[\{0,1\} \cup Q]$ is a set of cluster functions
on $\mathcal{A}$.
\end{framed}
\end{center}
Of course, additional functions of weight $k$ may be constructed by taking
products of functions of lower weight.

\begin{figure}
\begin{center}
\begin{tikzpicture}
\node[factNode,label=0:$x_1$] (1) at (0:1.5) {};
\node[factNode,label=36:$\tfrac{1}{x_3}$] (2) at (36:1.5) {};
\node[factNode,label=2*36:$x_5$] (3) at (2*36:1.5) {};
\node[factNode,label=3*36:$\tfrac{1}{x_2}$] (4) at (3*36:1.5) {};
\node[factNode,label=4*36:$x_4$] (5) at (4*36:1.5) {};
\node[factNode,label=5*36:$\tfrac{1}{x_1}$] (6) at (5*36:1.5) {};
\node[factNode,label=6*36:$x_3$] (7) at (6*36:1.5) {};
\node[factNode,label=7*36:$\tfrac{1}{x_5}$] (8) at (7*36:1.5) {};
\node[factNode,label=8*36:$x_2$] (9) at (8*36:1.5) {};
\node[factNode,label=9*36:$\tfrac{1}{x_4}$] (10) at (9*36:1.5) {};
%\draw[-,line width=1.2pt] (1.center) \foreach \i in {2,...,10}{-- (\i.center)}	-- cycle;
\draw[-,line width=1.2pt] (0,0) circle (1.5);
\end{tikzpicture}
\end{center}
\caption{The factorization graph for $A_2$. Each vertex is one of the
10 cluster coordinates on the $A_2$ cluster algebra (see eq.~(\ref{eq:A2coordinates})),
and two vertices
$x_i, x_j$ are connected by an edge if $x_i{-}x_j$ factors into
a product of cluster coordinates.  Each of the 10 pairs of connected
vertices, for example $\{1/x_2, x_5\}$, is a 2-clique.}
\label{fig:A2factorization}
\end{figure}
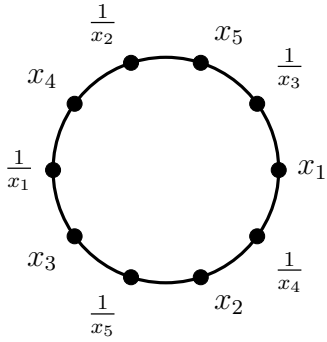

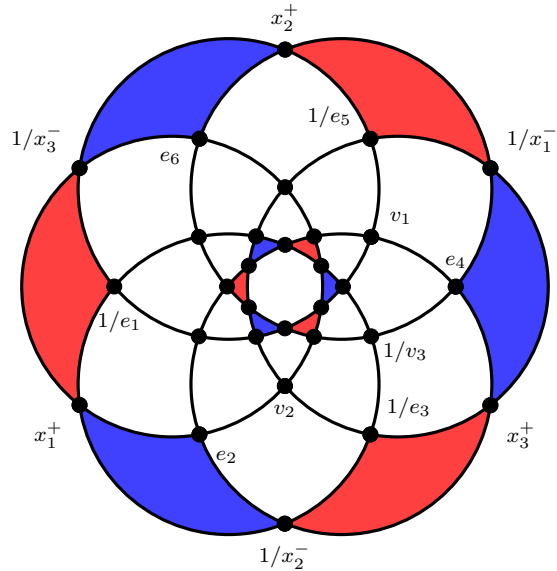
\begin{figure}
\begin{center}
\begin{tikzpicture}[remember picture,font=\scriptsize]
% A path that follows the edges of the current page
\def\ro{5.2}
\tikzstyle{reverseclip}=[insert path={(-40:\ro) --
  (-140:\ro) --
  (-220:\ro) --
  (-320:\ro) --
  cycle}
]
\def\ri{1.5}
\foreach\i in {1,3,5} {
	\begin{scope}
 		\draw [clip,draw=none](60*\i-60:\ri) circle (2cm) [reverseclip];
 		\draw [clip,draw=none](60*\i+60:\ri) circle (2cm) [reverseclip];
 		\fill[red2] (60*\i:\ri) circle (2cm);
	\end{scope}
	\begin{scope}
		\draw [clip,draw=none,name path global/.expanded = circle-\i](60*\i-60:\ri) circle (2cm);
		\draw [clip,draw=none](60*\i:\ri) circle (2cm) [reverseclip];
		\fill[blue2] (60*\i+60:\ri) circle (2cm);
	\end{scope}
}\foreach\i in {2,4,6} {
	\begin{scope}
 		\draw [clip,draw=none](60*\i-60:\ri) circle (2cm) [reverseclip];
 		\draw [clip,draw=none](60*\i+60:\ri) circle (2cm) [reverseclip];
 		\fill[blue2] (60*\i:\ri) circle (2cm);
	\end{scope}
	\begin{scope}
		\draw [clip,draw=none,name path global/.expanded = circle-\i](60*\i-60:\ri) circle (2cm);
		\draw [clip,draw=none](60*\i:\ri) circle (2cm) [reverseclip];
		\fill[red2] (60*\i+60:\ri) circle (2cm);
	\end{scope}
}
\foreach\i in {1,...,6}{
  \draw[draw=black,line width=1.2pt] (60*\i:\ri) circle (2cm);
}
\foreach \i in {1,...,5}{
	\pgfmathtruncatemacro{\k}{\i+1}	
	\foreach \j in {\k,...,6}{
		\node[name intersections={of={circle-\i} and {circle-\j}},factNode] at (intersection-1) {};
		\node[name intersections={of={circle-\i} and {circle-\j}},factNode] at (intersection-2) {};
	}
}

\node[name intersections={of={circle-1} and {circle-2}},factNode,label=45:$1/x_1^-$] at (intersection-1) {};
\node[name intersections={of={circle-2} and {circle-3}},factNode,label=90:$x_2^+$] at (intersection-1) {};
\node[name intersections={of={circle-3} and {circle-4}},factNode,label=135:$1/x_3^-$] at (intersection-1) {};
\node[name intersections={of={circle-4} and {circle-5}},factNode,label=225:$x_1^+$] at (intersection-2) {};
\node[name intersections={of={circle-5} and {circle-6}},factNode,label=270:$1/x_2^-$] at (intersection-2) {};
\node[name intersections={of={circle-6} and {circle-1}},factNode,label=315:$x_3^+$] at (intersection-2) {};

\node[name intersections={of={circle-1} and {circle-3}},factNode,label=175:$1/e_5$] at (intersection-1) {};
\node[name intersections={of={circle-2} and {circle-4}},factNode,label=200:$e_6$] at (intersection-1) {};
%\node[name intersections={of={circle-3} and {circle-5}},factNode,label=270:$1/e_1$] at (intersection-1) {};
\node[label=$1/e_1$] at (-2.2,-0.9) {};
\node[name intersections={of={circle-4} and {circle-6}},factNode,label=315:$e_2$] at (intersection-2) {};
\node[name intersections={of={circle-5} and {circle-1}},factNode,label=45:$1/e_3$] at (intersection-2) {};
\node[name intersections={of={circle-6} and {circle-2}},factNode,label=$e_4$] at (intersection-1) {};

\node[name intersections={of={circle-4} and {circle-1}},factNode,label=270:$v_2$] at (intersection-2) {};
%\node[name intersections={of={circle-5} and {circle-2}},factNode,label=300:$1/v_3$] at (intersection-1) {};
\node[label=$1/v_3$] at (1.6,-1.3) {};
\node[name intersections={of={circle-6} and {circle-3}},factNode,label=15:$v_1$] at (intersection-1) {};

\end{tikzpicture}
\end{center}
\caption{The factorization graph for $A_3$.  Each
vertex represents
one of the 30 cluster coordinates on the $A_3$ cluster algebra,
although to avoid clutter only 15 of the coordinates are labeled;
the other 15 are reciprocals of the ones shown.
Two vertices $x_i, x_j$ are connected by an edge
if $x_i{-}x_j \in M_{A_3}$.
The six circles each pass
through 10 vertices and indicate an $A_2$ subalgebra,
as shown in figure~\ref{fig:A2factorization}.  There are 12 subgraphs
with the topology of a triangle, 6 around the outer edge and 6 around
the inner edge; these are the 12 3-cliques.}
\label{fig:A3factorization}
\end{figure}

It may be helpful to visualize sets of cluster coordinates satisfying
the required property with the assistance of what we call a
{\bf factorization graph}.  For a given algebra $\mathcal{A}$, the
factorization graph contains one vertex for each cluster coordinate
$x \in~\hspace{-4pt}\mathcal{X}_\mathcal{A}$
and two vertices $x_i, x_j$ are connected if
$x_i - x_j \in M_{\mathcal{A}}$. The factorization graphs for
the $A_2$ and $A_3$ cluster algebras are shown in figures~\ref{fig:A2factorization} and~\ref{fig:A3factorization}.

In mathematics,
a complete subgraph (that is, a collection of vertices such that each
pair is connected by an edge) is known as a {\bf clique} (or an
$n$-clique, if it has $n$ vertices). It is evident from figures~\ref{fig:A2factorization}
and~\ref{fig:A3factorization} that $A_2$ has 10 2-cliques and no higher
cliques, while $A_3$ has 60 2-cliques, 12 3-cliques, and no higher cliques. Also note that the $A_3$ factorization graph is composed of 6 intersecting copies of the $A_2$ factorization graph. Therefore we can rephrase the conclusion boxed above by saying that
\begin{framed}
\noindent
\textbf{Cliques give cluster functions.}
\smallskip

\noindent
If $-Q \subseteq  \mathcal{X}_{\mathcal{A}}$ is a clique
of the factorization graph of $\mathcal{A}$, then
$G_k[\{0,1\} \cup Q]$ is a set of cluster functions
on $\mathcal{A}$.
\end{framed}

Since ordering will play a crucial role in what follows,
this is the perfect opportunity for us to note the convenient
fact that if $Q \subseteq
- \mathcal{X}_\mathcal{A}$ splits over $\mathcal{X}_\mathcal{A}$, then
there is a natural ordering on $Q$.  Recalling that cluster coordinates
are positive-valued everywhere in the interior of the positive domain,
possibly taking value 0 or $+\infty$ only on the boundary of that domain,
it is evident that for every pair $q_i \ne q_j \in Q$,
the difference
$q_i - q_j \in M_\mathcal{A}$ takes uniform
sign inside the positive domain.
Therefore, for each pair either $q_i < q_j$ or vice versa,
so the natural ordering on $Q$ is simply the true numerical
order $q_1<q_2< \cdots < q_n$ of these coordinates in the positive domain.
It will be convenient to choose the ordering on the set $\{0,1\} \cup Q$
to be $0, 1, q_1, \ldots, q_n$, even though this is not the true numerical
ordering of these quantities (since the $q$'s are negative in the positive
domain).

\subsection{Bases of Cluster Functions}

So far, we have seen that elements of the set $G_k[\st{0,1} \cup Q]$ are cluster functions on $\mathcal{A}$, i.e. have symbols which can be written in the symbol alphabet $\mathcal{X}_{\mathcal{A}}$ of cluster coordinates on $\mathcal{A}$, whenever $-Q$ is a clique for the factorization graph of $\mathcal{A}$. We now want a basis for $\mathcal{A}_\bullet(A_n)$, the space of cluster functions on $A_n$. Let's first consider a simple case. Suppose we are only interested in the space of polylogarithms with a fixed last argument:
\begin{equation}
	\mathcal{G}[S;z] = \bigoplus_{k=1}^\infty \; \spa G_k[S;z]\quad \text{ where }\quad 	G_k[S;z] = \st{G(s_1, \dots, s_k; z) \, : \, s_i \in S}
	\label{eq:fixed_last_arg_gons}
\end{equation}
where ``span'' denotes the vector space of $\mathbb{Q}$-linear combinations of the indicated functions.  We use the notation $\mathcal{G}$ to
carefully distinguish $G_k[S;z]$, which is a \emph{set} of weight-$k$ functions, from
$\mathcal{G}[S;z]$, which is a \emph{vector space} of functions of
any weight.

Although $\mathcal{G}[S;z]$ is a vector space, because of eq.~\eqref{eq:shuffle_identity} it is more useful to consider it as a \textbf{shuffle algebra}.
When dealing with such functions, it is more natural not to look for a vector space basis, but rather to find a minimal generating set for the algebra, such that each element of $\mathcal{G}[S;z]$ has a unique expression as a linear combination of \textit{products of} elements of the minimal generating set.

For this, we use Radford's Theorem (see~\cite{Radford}), which provides a minimal generating set for any free shuffle algebra in terms of \textbf{Lyndon words}. A Lyndon word of length $k$ on an ordered set $S$ is a sequence of $k$ elements of $S$ which is strictly smaller than all of its cyclic permutations  with respect to the lexicographic order of $S^k$. (Several explicit examples will be presented in section~\ref{sec:A3_Hedgehog_Basis}.) Let $\text{Lyndon}_k(S)$ denote Lyndon words of length $k$ on $S$. It is a consequence of Radford's Theorem that
$\mathcal{G}[S;z]$ has a minimal generating set
\begin{equation}
 \bigcup_{k \in \mathbb{N}} \, G_k[\text{Lyndon}_k(S); z].
\label{eq:radford_theorem}
\end{equation}

How can we use~\eqref{eq:radford_theorem} to generate cluster functions? The answer to this question\footnote{We are grateful to J.~Drummond for carefully explaining the application of Brown's results to the construction of functional bases.} is provided by Brown's extensive study of polylogarithms on the moduli spaces $\mathfrak{M}_{0,n+3}$\footnote{This is the space of configurations of $n+3$ distinct, ordered points on the Riemann sphere.} in~\cite{Brown} (see also~\cite{Bogner:2014mha} for some applications).
The results of~\cite{Brown} were presented in various useful coordinate systems on $\mathfrak{M}_{0,n+3}$. One key result was that (essentially) the space of Goncharov polylogarithms on $\mathfrak{M}_{0,n+3}$ is the tensor product of $n$ spaces of polylogarithms with fixed last arguments in a certain ordered set of variables $S$. The analysis in the previous subsection has revealed that choosing $S$ to be a clique $Q$ along with $\st{0,1}$ makes manifest the $A_n$ cluster structure of these functions. And, by eq.~\eqref{eq:radford_theorem}, we have a generating set for each of those $n$ spaces of polylogarithms. Combining these observations we arrive at:
\begin{framed}
	\noindent
	\textbf{For $A_n$, each $n$-clique gives a generating set for \textit{all} cluster functions.}

	\smallskip
\noindent
	If $-Q \subseteq \mathcal{X}_{A_n}$ is an $n$-clique of the factorization graph of the $A_n$ cluster algebra with an ordering $Q = \st{q_1 < q_2 < \cdots  < q_n}$, then
	\begin{equation}
		\bigcup_{k \in \mathbb{N}} \bigcup_{i=1}^n \, G_k[\text{Lyndon}_k\st{0,1,q_1, \dots, q_{i-1}}; q_i]
		\label{eq:hedgehog_basis}
	\end{equation}
	is a minimal generating set for $\mathcal{A}_\bullet(A_n)$, the space of cluster functions on $A_n$.
A vector space basis for $\mathcal{A}_\bullet(A_n)$ is given by
all possible products of elements of the set~\eqref{eq:hedgehog_basis}.
\end{framed}

We call the basis generated by eq.~\eqref{eq:hedgehog_basis} a \textbf{Hedgehog Basis} for reasons that will become clear in the next section. A very nice feature of this basis is that, thanks to the natural ordering
$q_1 < q_2 < \cdots <q_k<0$ on the set $Q$ discussed above,
it is manifest from eq.~(\ref{eq:Gdef}) that
each $G$ function in eq.~(\ref{eq:hedgehog_basis}) is free of branch cuts everywhere
in the interior of the positive domain, with possible branch cuts only on its
boundary --- with one important exception that we should note.
The exception is that at weight 1, instead of $G(0;q_i)$ we should use
the function
\begin{equation}
\label{eq:carefullog}
G(0;-q_i) = \log(-q_i).
\end{equation}

The feature of being free of branch cuts
in the positive domain is a necessary feature for these functions to be useful
in describing scattering amplitudes,
but the analytic constraints on amplitudes
are far stronger still: they must be singularity-free everywhere
inside the larger Euclidean domain, with branch points allowed only on
boundaries corresponding to multi-particle production thresholds.  It is an outstanding
problem of great importance to find an explicit basis for the subspace
of cluster functions spanned by functions satisfying these tighter
analytic constraints.

\subsection{The Hedgehog Theorem for $A_n$}
\label{sec:hedgehogtheorem}

We have now reduced the problem of finding a basis for cluster functions on $A_n$ to that of finding cliques $Q$ of size $n$. In this section we show that
there are precisely two such cliques for each $A_{n-1}$
subalgebra of $A_n$.  This correspondence can be visualized, at the level
of the exchange graph, by collections of cluster variables that we
call hedgehogs.

Let us start by defining hedgehogs. Suppose $\mathcal{A}$ is a cluster algebra of rank $r$ and $\mathcal{B}$ is a subalgebra of rank $r-1$. The exchange graph of $\mathcal{A}$ is an $r$-regular graph (each vertex has valence $r$), and the exchange graph for $\mathcal{B}$ is an embedded $(r-1)$-regular subgraph. Therefore, each vertex of $\mathcal{B}$ is incident to $r-1$ edges leading to other vertices of $\mathcal{B}$ and to one edge leading to a vertex of $\mathcal{A}\setminus \mathcal{B}$. In other words, each vertex of $\mathcal{B}$ has an edge which goes ``out of'' $\mathcal{B}$ and ``into'' $\mathcal{A}\setminus \mathcal{B}$. Recall from section~\ref{sec:cluster_algebras_review} that a directed edge of the exchange graph can be associated with a cluster coordinate $x \in \mathcal{X}_\mathcal{A}$. Let the \textbf{hedgehog} $\mathcal{X}(\mathcal{A}, \mathcal{B}) \subseteq \mathcal{X}_{\mathcal{A}}$ be the set of cluster coordinates associated to the edges going out of $\mathcal{B}$ into $\mathcal{A} \setminus \mathcal{B}$.

Example hedgehogs for $\mathcal{X}(A_2, A_1)$ and $\mathcal{X}(A_3, A_2)$ are shown in figures~\ref{fig:A2_A1_hedgehog} and~\ref{fig:A3_A2_hedgehog} respectively. As can be seen from the pictures, the edge variables in the set $\mathcal{X}(\mathcal{A}, \mathcal{B})$ radiate outwards --- just like the spines of a hedgehog. We might also consider the set of cluster coordinates associated to
inward directed edges, which just gives the ``anti-hedgehog''
\begin{equation}
\mathcal{X}^{-1}(\mathcal{A}, \mathcal{B}) = \{ 1/x : x \in \mathcal{X}(\mathcal{A}, \mathcal{B})\}.
\end{equation}
We are now in a position to state the main result of this paper:

\smallskip

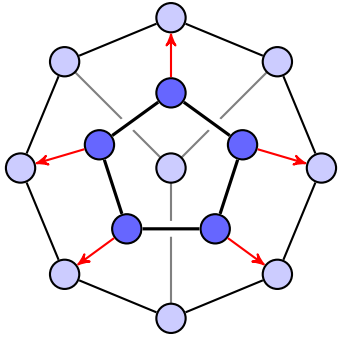
\begin{figure}
        \center
\begin{tikzpicture}[->,>=stealth',
thick,
  cross line/.style={preaction={draw=white, -,line width=6pt}}]
        
  \draw[-,black!50] (90+45:2) -- (0:0);
        \draw[-,black!50] (90-45:2) -- (0:0);
        \draw[-,black!50] (90+180:2) -- (0:0);

        \foreach \i in {1,...,5}{
                \node[blueNode,fill=blue!60] (p\i) at (90+\i*360/5:1) {};
        }
        \draw[-,cross line, very thick] (p1) -- (p2) -- (p3) -- (p4) -- (p5) -- (p1);
        
        \foreach \i in {1,...,8}{
                \node[blueNode] (o\i) at (90+\i*360/8:2) {};
        }
        \draw[-] (o1) -- (o2) -- (o3) -- (o4) -- (o5) -- (o6) -- (o7) -- (o8) -- (o1);

        \node[blueNode] (c) at (0:0) {};

        \draw[->,red,  thick] (p5) -- (o8);
        \draw[->,red,  thick] (p1) -- (o2);
        \draw[->,red,  thick] (p2) -- (o3);
        \draw[->,red,  thick] (p3) -- (o5);
        \draw[->,red,  thick] (p4) -- (o6);
\end{tikzpicture}

\caption{The $A_3$ algebra has six distinct hedgehogs (and
six anti-hedgehogs).  This figure shows
the exchange graph for $A_3$,
with one
of its six pentagonal $A_2$ subalgebras highlighted.
The ``spines'' of this
$\mathcal{X}(A_3,A_2)$
hedgehog are the red edges connecting this $A_2$ to the rest of
$A_3$.  Specifically, this $\mathcal{X}(A_3,A_2)$ is the set of 3
$\mathcal{X}_{A_3}$
cluster coordinates associated to these 5 outward directed red edges.}
\label{fig:A3_A2_hedgehog}
\end{figure}

\begin{framed}
\noindent
\textbf{The Hedgehog Theorem for $A_n$: Hedgehogs are $n$-Cliques}
\smallskip

\noindent
Let $\mathcal{X}(A_n,A_{n-1})$ be any hedgehog (or anti-hedgehog).
Then $Q = - \mathcal{X}(A_n,A_{n-1})$
is an $n$-clique of the factorization graph for $A_n$. In
particular, eq.~\eqref{eq:hedgehog_basis} generates a basis for the set of all
cluster functions on $A_n$.
\end{framed}

The details of the
proof of this theorem are presented in appendix~\ref{app:hedgehog},
using the machinery of triangulations reviewed in appendix~\ref{app:triangulations}.
Here we will be content to use the notation of the latter appendix
to provide explicit formulas for all $A_{n-1}$ hedgehogs of $A_n$, and to
check that they are cliques.

Let us note that the symbol alphabet of the cluster
functions generated by eq.~(\ref{eq:hedgehog_basis}),
which consists of letters of the form $q_i$, $1 - q_i$, or
$q_i - q_i$,
has exactly the same form as that of the polylogarithm
functions studied by Brown~\cite{Brown} in what he calls
simplicial
coordinates, $t_i$.  We are therefore able to conclude that
the two sets of functions can be related
to each other by the identification $t_i = - q_i = x_i$ between
simplicial coordinates $t_i$ on $\mathfrak{M}_{0,n+3}$
and the cluster coordinates $x_i$ of any hedgehog
$\mathcal{X}(A_n,A_{n-1})$
or anti-hedgehog $\mathcal{X}^{-1}(A_n,A_{n-1})$.

As reviewed in appendix~\ref{app:triangulations},
there are precisely $2\binom{n+3}{4}$ cluster variables on
$A_n$ (counting $x$ and $1/x$ separately); half of these can be enumerated
explicitly as cross-ratios
\begin{equation}
	r(i,j,k,\ell) = \frac{\braket{ij}\braket{k\ell}}{\braket{jk}\braket{i\ell}}, \qquad 1 \le i < j < k < \ell \le n + 3
	\label{eq:cross_ratios}
\end{equation}
of $n+3$ points in $\mathbb{P}^1$, while the other half are their
reciprocals $1/r(i,j,k,\ell) = r(j,k,\ell,i)$.
The $\Gr(2,n)$ Pl\"ucker coordinates $\ket{ij}$ used here may be related,
in the case $n=3$, to the $\Gr(4,n)$ coordinates used in section 2 by
\begin{equation}
\label{eq:pluckerrelation}
\ket{ij} = \frac{1}{4!} \epsilon_{ijk\ell mn} \ket{k\ell mn}.
\end{equation}

The $A_n$ cluster algebra has $n+3$ subalgebras of type $A_{n-1}$, so
there are $n+3$ hedgehogs.
In appendix~\ref{app:hedgehog} we show that these hedgehogs are given by
sets of the form
\begin{equation}
	\st{r(k, k+1, k+2, i) \, : \, i \not\in\st{k, k+1, k+2}}
	\label{eq:cliques_for_An}
\end{equation}
where $k+1$ and $k+2$ are taken mod $n$.
It is easy to verify that these are $n$-cliques by taking
two variables $r(k,k+1,k+2,i)$, $r(k,k+1,k+2, j)$ in this hedgehog and looking
at their difference,
\begin{equation}
\begin{aligned}
&r(k,k+1,k+2,i)-r(k,k+1,k+2,j)\\
\ &=\ 
\frac{\braket{k(k+1)}}{\braket{(k+1)(k+2)}\braket{ki}\braket{kj}}\left(\braket{(k+2)i}\braket{kj}-\braket{(k+2)j}\braket{ki}\right)\\
&= \frac{\braket{k(k+1)}}{\braket{(k+1)(k+2)}\braket{ki}\braket{kj}}\braket{k(k+2)}\braket{ji}\\
&= r(k,k+1,k+2,i)r(k,k+2,i,j) \in M_{A_n}
\end{aligned}
\label{eq:clique_different_computation}
\end{equation}
where the second equality is from a Pl\"ucker relation.

To summarize, for the $A_n$ cluster algebras, hedgehogs are cliques of size $n$.
There are $n+3$ hedgehogs, and $n+3$ anti-hedgehogs, related by the dihedral
symmetry
of the $n+3$-gon. This provides, via eq.~\eqref{eq:hedgehog_basis} and the Hedgehog Theorem,
$2(n+3)$ distinct, but equivalent, bases for cluster functions
on $A_n$.

\subsection{Comments on Other Algebras}
\label{sec:otheralgebras}

Our problem was to write down a basis of cluster functions on a cluster
algebra $\mathcal{A}$, and for $\mathcal{A} = A_n$ we have found that
eq.~\eqref{eq:hedgehog_basis} gives such a basis whenever $-Q$ is a hedgehog (or
anti-hedgehog) in $A_n$.
The algebras of most relevance to SYM theory, however,
are the $\Gr(4,n)$ algebras
(see~\cite{Golden:2013xva}).
Happily the one overlapping case $A_3 = \Gr(2,6) = \Gr(4,6)$ underlies the structure
of 6-particle scattering amplitudes.  We present
an application of our results to this case in the following
section.

For more general algebras $\mathcal{A}$, the definition of hedgehog
given above still makes sense, but it doesn't appear
to be useful.  In particular, it is straightforward to
check, for example, that there is no
$A_3 \subset D_4$, nor
$A_5 \subset E_6$, such that $\mathcal{X}(D_4,A_3)$ or $\mathcal{X}(E_6,A_5)$ are cliques.
For such hedgehogs, a set of functions of the type shown
in eqs.~\eqref{eq:Gkdef} or~\eqref{eq:hedgehog_basis} are still perfectly
fine sets of polylogarithm functions,
but they are not cluster functions: their symbols contain non-cluster
coordinates as entries.

One could, of course, look at smaller hedgehogs, associated
to $A_n \subset \mathcal{A}$ subalgebras, which are known to be cliques due to
the Hedgehog Theorem.  For example, $D_4$ has 12
distinct $A_3$ subalgebras, each of which has 6 $A_2$ subalgebras,
so in all there exist 72 $Q(A_3,A_2)$ hedgehogs
sitting inside $D_4$.  However it is easy to check that no individual
hedgehog furnishes
enough functions to provide a basis for all
cluster functions on $D_4$.
We know this because we can compare with the dimension of the spanning sets for
weight $\le 3$ described in eq.~\eqref{eq:trivialclusterfunctions}.
The same comment holds for $E_6$, which has seven $A_5$ subalgebras,
each of which in turn has eight $A_4$ subalgebras, for a total of
56 $\mathcal{X}(A_5, A_4)$ hedgehogs.
On the other hand, at least for the $D_4$ case we have checked that
the union of all cluster functions over these various hedgehogs provides
a vastly overcomplete set of cluster functions at weight $\le 3$, but we do not have
a collection of hedgehogs which exactly spans to provide a basis.
It may be that, just as eq.~\eqref{eq:hedgehog_basis} gives a basis for cluster functions on $A_n \cong \Gr(2,n)$ by gluing together sets of the form~\eqref{eq:radford_theorem} in a certain pattern, some different pattern of gluing might work for other algebras including the cases $\Gr(4,n)$ of relevance to scattering amplitudes.

\section{An Application to the 2-loop 6-particle NMHV Amplitude}

All evidence available to date
(including~\cite{DelDuca:2009au,DelDuca:2010zg,Goncharov:2010jf,Dixon:2011nj,Golden:2013xva,Dixon:2013eka,Dixon:2014voa,Dixon:2014iba})
supports the hypothesis that all
6-particle scattering amplitudes in SYM theory can be expressed
in terms of cluster functions on the $A_3$ cluster algebra.
As an application of the Hedgehog Theorem, we discuss
in this section
how to express the 2-loop 6-particle NMHV amplitude in a hedgehog basis.
This amplitude was originally computed in~\cite{Dixon:2011nj}
and written (see eq.~(2.27) of that paper)
as $[12345] (V + \tilde{V}) + {\rm cyclic}$,
where $[12345]$ is an $R$-invariant and $X \equiv 8 (V + \tilde{V})$
is a weight-4
polylogarithm function.
The exercise of rewriting $X$ in a hedgehog basis
has some practical benefit in that it produces a formula
which is notably shorter than results previously available in the literature.
But from our perspective a greater benefit of working
with a hedgehog basis is that it makes some of the cluster
structure of the amplitude manifest.

To highlight this point, let us note that
in the presentation of~\cite{Dixon:2011nj}, the amplitude $X$ is written
as a linear combination of
various generalized polylogarithm functions whose symbols may be written in the
10-letter alphabet
\begin{equation}
	\st{u,v,w, 1-u, 1-v, 1-w, y_u, y_v, y_w, 1-y_u y_v y_w}.
	\label{eq:y_alphabet}
\end{equation}
The relation between these variables and ours may be read off
from eq.~\eqref{eq:A3coords}.
The tenth letter $1-y_uy_vy_w$
is not ``clustery'' -- that is, it cannot be expressed
as a product of $A_3$ cluster coordinates, so it should
never appear in the symbol of anything we would call a cluster
function.
Indeed the full amplitude (like all 6-particle amplitudes)
has the property that
when all of the individual contributing polylogarithm
functions are added up, this tenth letter cancels out of the symbol of the full amplitude.
This is suggestive: if all these terms cancel out in the end, it seems
desirable to express the amplitude in such a way that they never
arise in the first place.
This is exactly what an $A_3$ hedgehog basis does.

\subsection{The Hedgehog Basis for $A_3$}
\label{sec:A3_Hedgehog_Basis}

Let's look at the hedgehog basis more concretely in the $A_3$ case.
For $n=3$, eq.~\eqref{eq:hedgehog_basis}
tells us that we can list basis elements for $A_3$ cluster
functions by enumerating
Lyndon words on the sets $\{0,1\}$, $\{0,1,q_1\}$, and $\{0,1,q_1,q_2\}$.

At weight 1 there are respectively 2, 3, 4 Lyndon words on these three
sets, which together provide the 9 weight-1 functions
in the $A_3$ basis:
\begin{align}
\{&G(0;-q_1),
G(1; q_1),
\nonumber
\\
&G(0;-q_2),
G(1; q_2),
G(q_1; q_2),
\label{eq:weight1}
\\
&G(0;-q_3),
G(1; q_3),
G(q_1; q_3),
G(q_2; q_3)\}.
\nonumber
\end{align}
(Here we recall that the three $G(0; z)$ functions are to be treated
as explained in eq.~\eqref{eq:carefullog}.)
At weight 2 there are respectively 1, 3, 6 Lyndon words on the three sets,
which together provide the 10 pure weight-2 functions:
\begin{align}
\{&G(0,1;q_1),
\nonumber
\\
&G(0,1;q_2),
G(0,q_1;q_2),
G(1,q_1;q_2),
\label{eq:weight2}
\\
&G(0,1;q_3),
G(0,q_1;q_3),
G(0,q_2;q_3),
G(1,q_1;q_3),
G(1,q_2;q_3),
G(q_1,q_2;q_3)\}.
\nonumber
\end{align}
An additional 45 functions of weight 2 may be obtained
by taking products of pairs of the weight-1 functions shown
in eq.~\eqref{eq:weight1}, so the total space of weight-2 functions
on $A_3$ has dimension 55.

It is a simple exercise to continue enumerating Lyndon words in this manner
to higher weight.
We find a total of
285 functions of weight 3 and 1351 functions of weight 4,
which is as far as we need to go for the purpose of
expressing the 2-loop amplitude $X$.
Symbols of functions in this hedgehog basis can be expressed in the 9-letter ``$q$'' alphabet
\begin{equation}
        \st{q_1, q_2, q_3, 1-q_1, 1-q_2, 1-q_3, q_1 - q_2, q_2 - q_3, q_1 - q_3}
        \label{eq:q_alphabet}
\end{equation}
where $-Q = \{-q_1, -q_2, -q_3\}$ is any 3-clique of the $A_3$ factorization
graph.

\subsection{Hedgehogs for $A_3$}

According to the
Hedgehog Theorem, cliques for $A_3$ are given precisely by hedgehogs (or anti-hedgehogs), which are in one-to-one correspondence with $A_2 \subset A_3$ subalgebras.
The hedgehogs for $A_3$ are triples of cluster coordinates associated with triangulations of a hexagon.
In terms of the variables defined in eq.~\eqref{eq:A3coords}, the six hedgehogs
are:
\begin{multline}
\text{Hedgehogs for } A_3 = \Big\{
\{1/x_3^-, 1/e_1, x_1^+\},
\{1/x_1^+, 1/e_2, x_2^-\},
\\
\{1/x_2^-, 1/e_3, x_3^+\},
\{1/x_3^+, 1/e_4, x_1^-\},
\{1/x_1^-, 1/e_5, x_2^+\},
\{1/x_2^+, 1/e_6, x_3^-\} \ \Big\}.
\label{eq:A3_hedgehogs}
\end{multline}
Each triple $\{x_1, x_2, x_3\}$ here
is listed in numerically increasing order in the positive domain (this can done consistently, see section~\ref{sec:good_arguments}), and we recall
that for each hedgehog $\{x_1, x_2, x_3\}$ there is a corresponding
anti-hedgehog $\{1/x_3, 1/x_2, 1/x_1\}$. These are highlighted in red and blue, respectively, in figure~\ref{fig:A3factorization}.
Altogether, the
twelve triples are related by dihedral transformations of the hexagon, or
equivalently by dihedral transformations of the six scattering particles.
Of course, the factorization and exchange graphs in figures~\ref{fig:A3factorization} and~\ref{fig:A3_A2_hedgehog} manifest this symmetry as well.

At this point, expressing the NMHV amplitude in such a hedgehog basis is an exercise in linear algebra. Since all 12 (anti-)hedgehogs give equivalent bases, we chose the basis from $\{1/x_3^+, e_3, x_2^-\}$ which provides the shortest representation of the pure weight 4 terms.  That means we use the basis explained in the previous subsection with the ordered set
\begin{equation}
\begin{aligned}
Q &= \{q_1, q_2, q_3\} = \{ - x_2^-, -e_3, -1/x_3^+ \}\\
&=\left\{
- (1 + x_2 + x_1 x_2) x_3, - \frac{(1+x_1)x_2 x_3}{1+x_3},
- \frac{x_1 x_2 x_3}{1 + x_3 + x_2 x_3}
\right\}\\
&=\left\{
- \sqrt{\frac{w}{u v y_u y_v y_w}},
- \sqrt{\frac{w(1-u)}{(1-w)u y_u y_w}},
- \sqrt{\frac{v w}{u y_u y_v y_w}}
\right\}.
\end{aligned}
\end{equation}
A simple calculation using the second line quickly reveals that the
difference of each pair lies in the multiplicative span of
the symbol letters shown in eq.~\eqref{eq:A3alphabet}, and also
that they are listed in increasing numerical order in the positive
domain.
These properties are less apparent from the third line.

Each of the first $9$ terms of the $``y"$ alphabet can be written as a product of elements of the $``q"$ alphabet so, by means of the symbol rule~\eqref{eq:symbolexpand}, the NMHV amplitude $X$ can be written in the ``$q$'' alphabet. Each element of the hedgehog basis can be expressed in the same alphabet and, because the symbol map is linear, the symbol of the amplitude can be written as a linear combination of the symbols of the basis vectors. To find the coefficients of this linear combination, it is convenient to work in the ambient $9^4$ dimensional space of length $4$ symbols in the ``$q$'' alphabet. The symbols of the hedgehog basis vectors, together with the amplitude, constitute 1352 linear combinations in this larger space with one linear relation. Calculating the null space of the $1352 \times 9^4$ matrix with the linear algebra library \texttt{SparseSuite} gives the appropriate linear combination. To summarize, the result of this calculation is
a particular linear combination of 376 elements of the
weight-4 hedgehog basis whose symbol matches that of the amplitude $X$ exactly.
To find a representation for the full amplitude we turn in the next
section to the problem of fixing terms of the form (transcendental numerical
coefficient) $\times$ (functions of weight less than four).

\subsection{Fixing Beyond-the-Symbol Terms}

If the symbols of two functions are equal, then the functions are equal, modulo ``beyond-the-symbol'' terms of lower weight. So this 376-term expression is the highest-weight part of the NMHV amplitude. A priori, we might expect up to 65 possible terms of lower weight. These include 55 weight-2 functions times $\zeta(2)$, 9 weight-1 functions times $\zeta(3)$, and one overall additive constant proportional to $\zeta(4)$. The coefficients of these 65 terms can be fixed by numerically evaluating the amplitude and our 376-term highest-weight expression at 65 random points in the positive domain and performing a row reduction. All the coefficients turn out to be rational numbers with small denominators.
Our final result\footnote{Our result is included as an ancillary file with the arXiv submission.} is a 416-term expression for the 2-loop, 6-particle NMHV amplitude $X$.
The validity of our ansatz, and solution,
for the lower-weight terms has been stringently tested by comparing our result
to the known expression at high precision for additional random kinematic
points\footnote{We are grateful to L.~J.~Dixon and A.~McLeod for kindly
providing an independent audit of our result in this manner.}.

\section{Outlook}

Hedgehog bases give a natural way to express 6-particle amplitudes, since they make manifest that these amplitudes have symbols which can be expressed in terms of $A_3$ cluster coordinates.
In practice, this may translate into more ``compact'' representations of amplitudes than might be otherwise achieved.
It should be stressed again that this the hedgehog basis is a true basis for cluster functions, with no functional or linear relations between its elements.

However, hedgehog bases are clearly not the ultimate solution for representing scattering amplitudes. The most important reason is that amplitudes satisfy a stringent analytic constraint on the possible locations of their branch points, which translates into a condition that allows only certain letters to appear in the first entry of their symbols.  For example, 6-particle amplitudes may only have the letters $\{u, v, w\}$ in the first entry of their symbols, whereas
all nine letters of the $A_3$ symbol alphabet appear as first entries
in the hedgehog basis.
It would be extremely interesting, as well as of great practical utility,
to see if there is a natural way to construct bases of cluster functions
manifesting this additional property.
It would also be very interesting, both mathematically and
physically, to find an appropriate extension of the Hedgehog Theorem to algebras
other than $A_n$.

\acknowledgments

We have benefitted from enormously valuable discussions with J.~Drummond,
correspondence with L.~Dixon, and collaboration on closely related topics
with J.~Golden and A.~Goncharov.
This work was supported by the US Department of Energy under
contract DE-SC0010010 (MS) and the DE-FG02-11ER41742 Early Career Award (AV),
by the Sloan Research Foundation (AV), and by Brown University UTRA Awards
(DP and AS).
MS and AV are also grateful to the CERN Theory Group for support during the
completion of this work.

\appendix

\section{Triangulations and $A_n$ Cluster Algebras}
\label{app:triangulations}

Here we review from~\cite{1054.17024} the fact that
in the special case case of $A_n$ cluster algebras there is a convenient alternative to representing clusters with quivers: each cluster can instead be associated with a triangulation of an $(n+3)$-sided polygon.  Beginning with a labeled $(n+3)$-gon, a triangulation is obtained by repeatedly adding non-crossing internal chords $\overline{ik}$ between nonadjacent vertices $i, k$ until no further chords can be added. There are always $n$ chords in a triangulation.  For example, the five chords in a particular triangulation of an octagon are shown here:
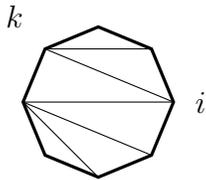
\begin{center}
\begin{tikzpicture}

	\foreach \i in {0,...,7}{
		\node[vertex] (v\i) at (\i*360/8:1) {};
	}
	\draw[polygonEdge] (0:1) \foreach \i in {1,...,7} {
		-- (\i*360/8:1)
	}--cycle;

	\draw[-] (v0) node[label={0:$i$}] {} -- (v3) node[label={135:$k$}] {};
	\draw[-] (v1) -- (v3);
	\draw[-] (v4) -- (v0);
	\draw[-] (v4) -- (v7);
	\draw[-] (v4) -- (v6);
\end{tikzpicture}
\end{center}
Cluster coordinates are associated with these chords.
Specifically, the chord $\overline{ik}$ shared between two triangles $ijk$ and $ikl$
is associated with the cluster variable\footnote{Here we use an inverse convention compared to~\cite{Scott} and parts of~\cite{Golden:2013xva}.}
\begin{equation}
r(j,k,\ell, i)=
\frac{1}{r(i,j,k,\ell)}:=\frac{\braket{jk}\braket{i\ell}}{\braket{ij}\braket{k\ell}},
\label{eq:2_bracket_cross_ratio}
\end{equation}
where $\braket{ij}$ denotes the Pl\"ucker coordinate of two points
$z_i, z_j$
in $\mathbb{P}^1$.
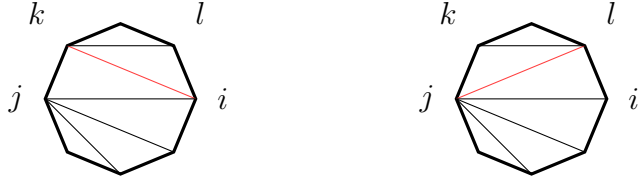
\begin{center}
\begin{tikzpicture}

	\foreach \i in {0,...,7}{
		\node[vertex] (v\i) at (\i*360/8:1) {};
	}
	\draw[polygonEdge] (0:1) \foreach \i in {1,...,7} {
		-- (\i*360/8:1)
	}--cycle;

	\draw[-,red2] (v0) -- (v3);
	\draw[-] (v1) node[label={45:$l$}] {} -- (v3) node[label={135:$k$}] {};
	\draw[-] (v4) node[label={180:$j$}] {} -- (v0) node[label={0:$i$}] {};
	\draw[-] (v4) -- (v7);
	\draw[-] (v4) -- (v6);
\end{tikzpicture}
\hspace{2cm}
\begin{tikzpicture}

	\foreach \i in {0,...,7}{
		\node[vertex] (v\i) at (\i*360/8:1) {};
	}
	\draw[polygonEdge] (0:1) \foreach \i in {1,...,7} {
		-- (\i*360/8:1)
	}--cycle;

	\draw[-,red2] (v1) {} -- (v4) {};
	\draw[-] (v1) node[label={45:$l$}] {} -- (v3) node[label={135:$k$}] {};
	\draw[-] (v4) node[label={180:$j$}] {} -- (v0) node[label={0:$i$}] {};
	\draw[-] (v4) -- (v7);
	\draw[-] (v4) -- (v6);
\end{tikzpicture}
\end{center}

In this representation, mutations are associated with chord-flips. To perform a chord-flip on $r(i,j,k,\ell)$, remove the chord $\overline{ik}$ and add the chord $\overline{jl}$. It is easy to see that the resulting variable $r(j,k,\ell,i)$ is indeed equal to $1/r(i,j,k,\ell)$; adjacent chords (those which lie on the same triangle) take the place of adjacent nodes in a quiver.

The added convenience of using triangulations over quivers comes from the fact that every triangulation is associated with a single cluster whose variables can be found explicitly via the formula above. Using this explicit formula, one can determine many useful facts about $A_n$: it's order is the Catalan number $C_{n+1}$, there are $2\binom{n+3}{4}$ cluster variables, and those variables are $r(i,j,k,\ell)$ for cyclically ordered $i,j,k,\ell$.

Triangulations also make it easy to enumerate and analyze subalgebras. Consider the case $n=5$. Clusters of $A_5$ correspond to triangulations of a labeled octagon. Selecting the vertices $1,3,4,7,$ and $8$, we can form a pentagon within our octagon:
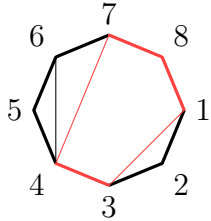
\begin{center}
\begin{tikzpicture}
	
	\foreach \i in {2,5,6}{
	\node[vertex,label={45+-\i*360/8:\i}] (v\i) at (45+-\i*360/8:1) {};
	}

	\foreach \i in {1,3,4,7,8}{
	\node[vertex,label={45+-\i*360/8:\i},red2] (v\i) at (45+-\i*360/8:1) {};
	}

	\draw[polygonEdge] (v7) -- (v6) -- (v5) -- (v4) (v3) -- (v2) -- (v1);
	\draw[-,red2] (v1) -- (v3) (v4) -- (v7);
	\draw[-,red2,polygonEdge] (v3) -- (v4) (v7) -- (v8) -- (v1);
	\draw[-] (v4) -- (v6);

\end{tikzpicture}
\end{center}
Some triangulations of the octagon contain all of the edges of the pentagon as chords $(\overline{13},\overline{34},\overline{47},\overline{48},\overline{81})$. The subtriangulation obtained by discarding everything outside the pentagon is associated with a cluster of $A_2$. By flipping only the chords lying strictly within the pentagon, we can obtain other $A_2$ clusters, until we have an entire $A_2$ subalgebra:
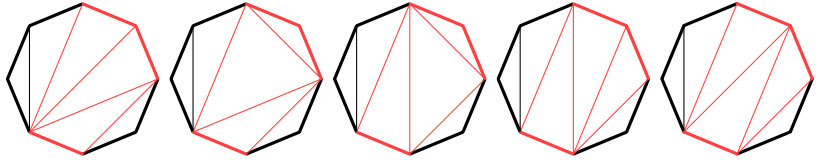
\begin{center}
\begin{tikzpicture}
	
	\foreach \i in {2,5,6}{
	\node[vertex] (v\i) at (45+-\i*360/8:1) {};
	}
	%\draw[polygonEdge] (0:1) \foreach \i in {1,...,7} {
	%	-- (\i*360/8:1)
	%}--cycle;

	\foreach \i in {1,3,4,7,8}{
	\node[vertex,red2] (v\i) at (45+-\i*360/8:1) {};
	}

	\draw[polygonEdge] (v7) -- (v6) -- (v5) -- (v4) (v3) -- (v2) -- (v1);
	\draw[-,red2] (v1) -- (v3) (v4) -- (v7);
	\draw[-,red2,polygonEdge] (v3) -- (v4) (v7) -- (v8) -- (v1);
	\draw[-] (v4) -- (v6);
	
	\draw[-,red2] (v1) -- (v4) (v4) -- (v8);
\end{tikzpicture}
\begin{tikzpicture}
	
	\foreach \i in {2,5,6}{
	\node[vertex] (v\i) at (45+-\i*360/8:1) {};
	}
	%\draw[polygonEdge] (0:1) \foreach \i in {1,...,7} {
	%	-- (\i*360/8:1)
	%}--cycle;

	\foreach \i in {1,3,4,7,8}{
	\node[vertex,red2] (v\i) at (45+-\i*360/8:1) {};
	}

	\draw[polygonEdge] (v7) -- (v6) -- (v5) -- (v4) (v3) -- (v2) -- (v1);
	\draw[-,red2] (v1) -- (v3) (v4) -- (v7);
	\draw[-,red2,polygonEdge] (v3) -- (v4) (v7) -- (v8) -- (v1);
	\draw[-] (v4) -- (v6);
	
	\draw[-,red2] (v1) -- (v4) (v7) -- (v1);

\end{tikzpicture}
\begin{tikzpicture}
	
	\foreach \i in {2,5,6}{
	\node[vertex] (v\i) at (45+-\i*360/8:1) {};
	}
	%\draw[polygonEdge] (0:1) \foreach \i in {1,...,7} {
	%	-- (\i*360/8:1)
	%}--cycle;

	\foreach \i in {1,3,4,7,8}{
	\node[vertex,red2] (v\i) at (45+-\i*360/8:1) {};
	}

	\draw[polygonEdge] (v7) -- (v6) -- (v5) -- (v4) (v3) -- (v2) -- (v1);
	\draw[-,red2] (v1) -- (v3) (v4) -- (v7);
	\draw[-,red2,polygonEdge] (v3) -- (v4) (v7) -- (v8) -- (v1);
	\draw[-] (v4) -- (v6);
	
	\draw[-,red2] (v7) -- (v3) (v7) -- (v1);

\end{tikzpicture}
\begin{tikzpicture}
	
	\foreach \i in {2,5,6}{
	\node[vertex] (v\i) at (45+-\i*360/8:1) {};
	}
	%\draw[polygonEdge] (0:1) \foreach \i in {1,...,7} {
	%	-- (\i*360/8:1)
	%}--cycle;

	\foreach \i in {1,3,4,7,8}{
	\node[vertex,red2] (v\i) at (45+-\i*360/8:1) {};
	}

	\draw[polygonEdge] (v7) -- (v6) -- (v5) -- (v4) (v3) -- (v2) -- (v1);
	\draw[-,red2] (v1) -- (v3) (v4) -- (v7);
	\draw[-,red2,polygonEdge] (v3) -- (v4) (v7) -- (v8) -- (v1);
	\draw[-] (v4) -- (v6);
	
	\draw[-,red2] (v7) -- (v3) (v8) -- (v3);

\end{tikzpicture}
\begin{tikzpicture}
	
	\foreach \i in {2,5,6}{
	\node[vertex] (v\i) at (45+-\i*360/8:1) {};
	}
	%\draw[polygonEdge] (0:1) \foreach \i in {1,...,7} {
	%	-- (\i*360/8:1)
	%}--cycle;

	\foreach \i in {1,3,4,7,8}{
	\node[vertex,red2] (v\i) at (45+-\i*360/8:1) {};
	}

	\draw[polygonEdge] (v7.center) -- (v6.center) -- (v5.center) -- (v4.center) (v3.center) -- (v2.center) -- (v1.center);
	\draw[-,red2] (v1) -- (v3) (v4) -- (v7);
	\draw[-,red2,polygonEdge] (v3.center) -- (v4.center) (v7.center) -- (v8.center) -- (v1.center);
	\draw[-] (v4) -- (v6);
	
	\draw[-,red2] (v8) -- (v4) (v3) -- (v8);

\end{tikzpicture}
\end{center}
By flipping the chords lying strictly outside the pentagon, or choosing a different pentagon to begin with, we can obtain different $A_2$ subalgebras. Note that the different subalgebras sharing the same pentagonal boundary must all have the same set of cluster variables; therefore, if we consider two subalgebras ``equivalent'' when they have the same variables; there are exactly $\binom{8}{5}=56$ nonequivalent $A_2$ subalgebras of $A_5$.

This generalizes nicely: the clusters of $A_m$ subalgebras of $A_n$ correspond to $(m+3)$-vertex subtriangulations of $(n+3)$-gon triangulations. Up to equivalence, there are $\binom{n+3}{m+3}$ of these.

\section{A Theorem on $1+\mathcal{X}$ Coordinates}
\label{app:theorem}

In this appendix we prove that for all cluster coordinates
$x$, the quantity $1 + x$ can be expressed as a product
of cluster coordinates on the same algebra.

	To be precise: Suppose $C$ is a cluster algebra of $A, D$, or $E$ type whose quivers are connected with more than one node.  Suppose $\mathcal{X}_C$ is its set of cluster coordinates. Then if $x_i \in \mathcal{X}_C$,
	\begin{equation}
		1 + x_i = \prod_{x_j \in \mathcal{X}} x_j^{n_j}
	\end{equation}
	for some $n_j \in \st{-1,0,1}$.

	The proof of this statement is straightforward.  Pick some quiver of $C$ that contains $x_i$. By connectedness, there exists some $x_k$ connected to it. One of the properties of an A,D or E-type cluster algebra is that $\abs{B_{ij}} \le 1$ for all $i,j$. In particular, $B_{ik} \in \st{-1,0,1}$. Recall the mutation rule for cluster coordinates:
\begin{equation}
\label{eq:mutationrule}
	x_k' = \mu_i (x_k)  = \begin{cases}
		x_i^{-1} & i = k\\
		x_k \Big(1 + x_i^{\sgn B_{ik}}\Big)^{B_{ik}} & i \neq k.
	\end{cases}
\end{equation}
If $B_{ik} = 1$, then
\begin{equation}
	x_k' = x_k\left(1+x_i^1\right)^1 \implies 1+ x_i = \frac{x_k'}{x_k}.
\end{equation}
Otherwise, $B_{ik} = -1$, in which case
\begin{equation}
	x_k' = x_k \left( 1+x_i^{-1} \right)^{-1} = x_k \left( \frac{x_i}{1+x_i} \right)  \implies 1+x_i = \frac{x_i x_k}{ x_k'}.
\end{equation}
Since $x_i$ and $x_k$ are connected, $B_{ik} \neq 0$, so this is exhaustive. Thus $1+x_i$ factors as a product of cluster coordinates.

Connectedness of a quiver is preserved by mutation, so if the initial quiver is connected, all quivers of an algebra are connected. Note also that the algebra $A_1$ as well as its derived algebras such as $A_1 \times A_2$ have disconnected quivers, hence expressions of the form $1+x$ do not necessarily factor in this case.

Let us note here that a sort of converse statement,
which has been stated and used for example
in~\cite{Golden:2013xva,Golden:2013lha,Golden:2014pua}, remains a conjecture:
If $a$ and
$b$ are two elements of the multiplicative span $M_{\mathcal{A}}$
for some cluster algebra $\mathcal{A}$, and if
$b = a + 1$, then precisely one element of the set $\{a, -1-a, -1-1/a\}$ is
a cluster coordinate.

\section{A Cluster Parameterization of 6-Particle Kinematics}
\label{app:kinematics}

We include here a parameterization of the positive domain of
6-particle scattering kinematics, in terms of momentum twistors,
that we have found useful:
\begin{equation}
Z = \left(\begin{matrix}
-1 & 0 & 0 & 0 & 1 & 1 + x_1 \\
0&1&0&0&1+x_2&1+x_2+x_1 x_2\\
0&0&-1&0&1+x_3+x_2 x_3&1+x_3 + x_2 x_3 + x_1 x_2 x_3\\
0&0&0&1&1&1
\end{matrix}
\right).
\end{equation}
It is easily checked that this lies in the positive domain
(that is, all minors $\ket{ijkl} > 0$ when $i<j<k<l$)
whenever $x_1,x_2,x_3 > 0$, and that when plugged into the last
column of eq.~\eqref{eq:A3coords}, it precisely reproduces the second
column.

\section{Hedgehogs are Cliques for $A_n$ Cluster Algebras}
\label{app:hedgehog}

Here we provide the details of the proof of the Hedgehog Theorem presented
in section~\ref{sec:hedgehogtheorem}.
First consider the case $n=2$. The general case can be reduced to the $n=2$ case, so it is worth doing in detail.

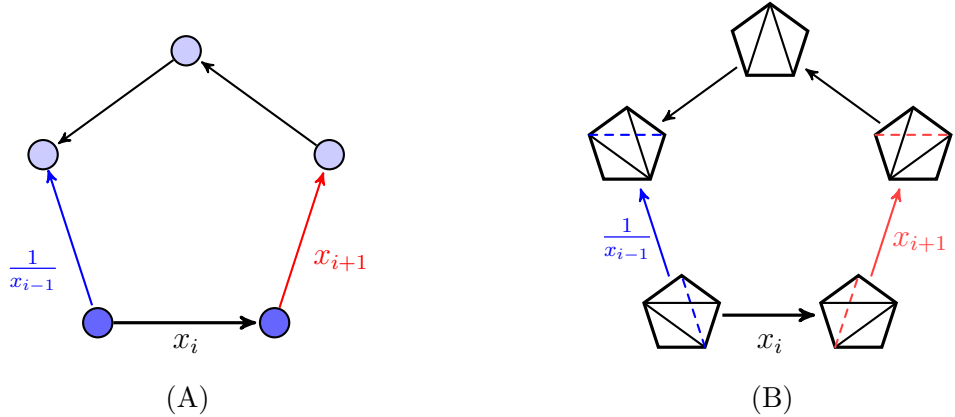
\begin{figure}
	\center
	\begin{subfigure}[b]{0.5\textwidth}\centering
	\begin{tikzpicture}[->,>=stealth',shorten >=1pt,auto,node distance=2cm,
  thick]
 		\node[blueNode] (1) at (18:2) {};
 		\node[blueNode] (2) at (72+18:2) {};
 		\node[blueNode] (3) at (144+18:2) {};
 		\node[blueNode,fill=blue!60] (4) at (216+18:2) {};
 		\node[blueNode,fill=blue!60] (5) at (288+18:2) {};
		\draw[->] (1) -- (2);
		\draw[->] (2) -- (3);
		\draw[blue,<-] (3) -- (4) node[midway, label=below left : $\frac{1}{x_{i-1}}$] {};
		\draw[->,very thick] (4) -- (5) node[midway, label=below: $x_{i}$] {};
		\draw[red, ->] (5) -- (1) node[midway, label=below right: $x_{i+1}$] {};
	\end{tikzpicture}
	\caption{}
	\label{fig:A2_A1_hedgehog}
\end{subfigure}%
	\begin{subfigure}[b]{0.5\textwidth}\centering
	\begin{tikzpicture}[->,>=stealth',shorten >=1pt,auto,node distance=2cm,
  thick]

	\foreach \j in {1,...,5}{
		\node[draw,circle,minimum size = 30pt,white] (\j) at (72*\j+18:2) {};
		\foreach \i in {0,...,5}{\node[vertex] (p\j\i) at ($(\j) + (18+\i*360/5:15pt)$) {};}
			\draw[-,polygonEdge] (p\j1.center) \foreach \i in {2,...,5}{-- (p\j\i.center)}	-- cycle;
	}

	\draw[-,line cap=round] (p13.center) -- (p11.center);
	\draw[-,line cap=round] (p14.center) -- (p11.center);

	\draw[-,line cap=round] (p24.center) -- (p22.center);
	\draw[-,line cap=round] (p24.center) -- (p21.center);
	\draw[-,line cap=round,dashed, blue] (p22.center) -- (p25.center);

	\draw[-,line cap=round] (p34.center) -- (p32.center);
	\draw[-,line cap=round, dashed, blue] (p34.center) -- (p31.center);
	\draw[-,line cap=round] (p32.center) -- (p35.center);

	\draw[-,line cap=round] (p43.center) -- (p45.center);
	\draw[-,line cap=round, dashed, red2] (p43.center) -- (p41.center);
	\draw[-,line cap=round] (p42.center) -- (p45.center);

	\draw[-,line cap=round] (p53.center) -- (p55.center);
	\draw[-,line cap=round] (p53.center) -- (p51.center);
	\draw[-,line cap=round, dashed, red2] (p52.center) -- (p55.center);

	\node[] at (72*3.5+18:2) {$x_i$};
	\node[blue] at (72*2.5+18:2) {$\frac{1}{x_{i-1}}$};
	\node[red2] at (72*4.5+18:2.1) {$x_{i+1}$};

	\draw[->,blue] (3) -- (2);
	\draw[->, very thick] (3) -- (4);
	\draw[->,red2] (4) -- (5);
	\draw[->] (5) -- (1);
	\draw[->] (1) -- (2);

\end{tikzpicture}
\caption{}
\label{fig:A2_A1_hedgehog_triangulation}
\end{subfigure}%
\caption{(A) The exchange graph for $A_2$, with the two vertices on the bottom row constituting an $A_1$ subalgebra.
The hedgehog $\mathcal{X}(A_2,A_1)$ contains the two cluster variables $1/x_{i-1}$, $x_{i+1}$ associated to the edges
emanating away from the subalgebra.  (B) The same exchange graph, but with each vertex showing the associated
pentagon triangulation.}
\end{figure}

As reviewed in section~\ref{sec:cluster_algebras_review},
the mutation relations for $\mathcal{A} \cong A_2$ give ten cluster coordinates $\st{x_1, \tfrac{1}{x_1}, \dots, x_5, \tfrac{1}{x_5}}$ related by
\begin{equation}
	x_{i+1} = \frac{1+x_i}{x_{i-1}}.
\end{equation}
For $\mathcal{B} \cong A_1$ let us choose the subalgebra with coordinates $\st{x_i, \tfrac{1}{x_i}}$. So the relevant hedgehog is $\mathcal{X}(\mathcal{A},\mathcal{B}) =  \st{x_{i+1}, \frac{1}{x_{i-1}}}$. Pictorially, the hedgehog consists of the red and blue edges in the exchange graph shown in figure~\ref{fig:A2_A1_hedgehog}.

This is a clique because
\begin{equation}
	x_{i+1} - \frac{1}{x_{i-1}} = \frac{1+x_i}{x_{i-1}} - \frac{1}{x_{i-1}} = \frac{x_i}{x_{i+1}} \in M_{\mathcal{A}}.
\end{equation}
Recasting this in terms of polygon triangulation is very illuminating. Recall that $A_2$ can be described in terms of pentagon triangulations.
The red dashed lines in figure~\ref{fig:A2_A1_hedgehog_triangulation} are the chords that change as the red edge is traversed, and similarly for the blue. What we have shown then is that the difference between the cluster coordinates for the red and blue edges can be written in terms of products of cluster coordinates.

Now for the general case consider a $(n+3)$-gon. Choose three adjacent vertices $k, k+1, k+2$ and draw the chord $\overline{k (k+2)}$. This chord separates a triangle from an $(n+2)$-gon. The variable associated with this chord depends only on the triangles containing it, and not on the rest of the triangulation. The triangle on one side of the chord will always have the vertices $k,k+1,k+2$. The other triangle will have vertices $k,k+2,i$, where $i$ is any of the $n$ remaining vertices:
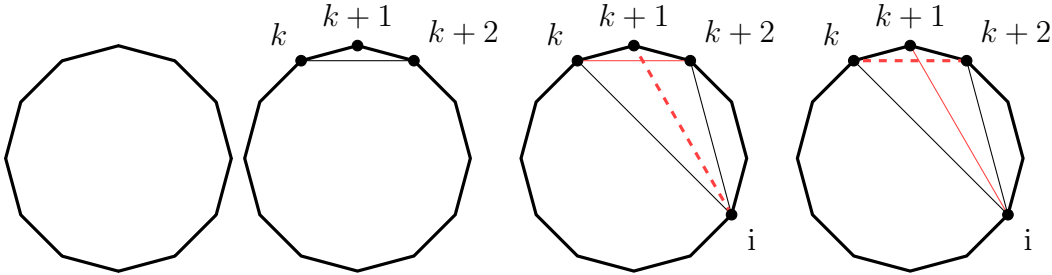
\begin{center}
\begin{tikzpicture}
	\foreach \i in {1,2,3,4,5,6,7,8,9,10,11,12}{
	\node[vertex] (v\i) at (0+-\i*360/12:1.5) {};
	}
	
	\draw[polygonEdge] (v12) -- (v11) -- (v10) -- (v9) -- (v8) -- (v7) -- (v6) -- (v5) -- (v4) -- (v3) -- (v2) -- (v1) -- (v12);
\end{tikzpicture}
\begin{tikzpicture}
	\foreach \i in {1,...,12}{
	\node[vertex] (v\i) at (0+-\i*360/12:1.5) {};
	}
	
	\draw[polygonEdge] (0:1.5) \foreach \i in {1,...,12} {
		-- (\i*360/12:1.5)
	}--cycle;
	
	\draw[-] (v10) -- (v8);
	
	\draw[-] (v8) node[label={-8*360/12:$k$},vertex2] {};
	\draw[-] (v9) node[label={-9*360/12:$k+1$},vertex2] {};
	\draw[-] (v10) node[label={-10*360/12:$k+2$},vertex2] {};
\end{tikzpicture}
\begin{tikzpicture}
	\foreach \i in {1,...,12}{
	\node[vertex] (v\i) at (-\i*360/12:1.5) {};
	}
	
	\draw[polygonEdge] (0:1.5) \foreach \i in {1,...,12} {
		-- (\i*360/12:1.5)
	}--cycle;
	
	\draw[-,red2] (v10) -- (v8);
	\draw[-] (v8) -- (v1) -- (v10);
	\draw[dashed,red2,very thick] (v1) -- (v9);

	\draw[-] (v8) node[label={-8*360/12:$k$},vertex2] {};
	\draw[-] (v9) node[label={-9*360/12:$k+1$},vertex2] {};
	\draw[-] (v10) node[label={-10*360/12:$k+2$},vertex2] {};
	
	\draw[-] (v1) node[label={-1*360/12:i},vertex2] {};
\end{tikzpicture}
\begin{tikzpicture}
	\foreach \i in {1,...,12}{
	\node[vertex] (v\i) at (-\i*360/12:1.5) {};
	}
	
	\draw[polygonEdge] (0:1.5) \foreach \i in {1,...,12} {
		-- (\i*360/12:1.5)
	}--cycle;
	
	\draw[dashed,red2,very thick] (v10) -- (v8);
	\draw[-] (v8) -- (v1) -- (v10);
	\draw[-,red2] (v1) -- (v9);
	
	\draw[-] (v8) node[label={-8*360/12:$k$},vertex2] {};
	\draw[-] (v9) node[label={-9*360/12:$k+1$},vertex2] {};
	\draw[-] (v10) node[label={-10*360/12:$k+2$},vertex2] {};
	
	\draw[-] (v1) node[label={-1*360/12:i},vertex2] {};
\end{tikzpicture}
\end{center}
Therefore, there are $n$ cluster variables that can be associated with $\overline{k(k+2)}$. These $n$ variables are given by
\begin{equation}
\{r(i,k,k+1,k+2):i\notin\{k,k+1,k+2\}\}.
\end{equation}
Consider the subalgebra $B\cong A_{n-1}$ associated with the $(n+2)$-gon that excludes vertex $(k+1)$. Any triangulation containing $\overline{k(k+2)}$ will contain a triangulation of this polygon, and hence will be associated with a cluster in $B$. However, flipping $\overline{k(k+2)}$ will yield a triangulation that is not in $B$; therefore, the set of cluster variables associated with $\overline{k(k+2)}$ is a hedgehog of $A_n$! Because there are $n+3$ choices for $k$, all hedgehogs can be so obtained, and all will be of cardinality $n$.

We can also obtain the anti-hedgehogs, which are associated with the result of any chord-flip of $\overline{k(k+2)}$:
\begin{equation}
 \{r(k,k+1,k+2,i):i\notin\{k,k+1,k+2\}\}.
\end{equation}

Take $x_i, x_j$ to be two arbitrary elements of the hedgehog, with $(i,j,k)$ cyclically ordered. Then the mutation $x_i\mapsto1/x_i$ corresponds to flipping $\overline{k(k+2)}$ to $\overline{(k+1)i}$ and $x_j\mapsto1/x_j$ corresponds to flipping $\overline{k(k+2)}$ to $\overline{(k+1)j}$. These are indicated in figure~\ref{fig:HedgeProof} in (C) and (A) respectively.
\begin{figure}[h]
\centering
\begin{subfigure}[b]{0.25\textwidth}\center
\begin{tikzpicture}
	\foreach \i in {1,...,12}{\node[vertex] (v\i) at (-\i*360/12:1.5) {};}
	
	\draw[-,grayOut] (v4.center) \foreach \i in {5,...,8}{-- (v\i.center)}	-- cycle;
	\draw[-,grayOut] (v4.center) \foreach \i in {4,3,2,1,12,11,10}{ -- (v\i)}	-- cycle;

	\draw[-] (v10) -- (v8);
	\draw[-] (v8) -- (v4) -- (v10);
	\draw[dashed,blue,very thick] (v4) -- (v9);
	
	\draw[-] (v8) node[label={-8*360/12:$k$},vertex2] {};
	\draw[-] (v9) node[label={-9*360/12:$k+1$},vertex2] {};
	\draw[-] (v10) node[label={-10*360/12:$k+2$},vertex2] {};
	
	\draw[-] (v1) node[label={-1*360/12:$i$},vertex2] {};
	\draw[-] (v4) node[label={-4*360/12:$j$},vertex2] {};

	\draw[-,polygonEdge] (v1.center) \foreach \i in {2,...,12}{-- (v\i.center)}	-- cycle;
\end{tikzpicture}
\caption{}
\end{subfigure}%
\begin{subfigure}[b]{0.25\textwidth}\centering
\begin{tikzpicture}
	\foreach \i in {1,...,12}{\node[vertex] (v\i) at (-\i*360/12:1.5) {};}
	
	\draw[-,grayOut] (v4.center) \foreach \i in {5,...,8}{-- (v\i.center)}	-- cycle;
	\draw[-,grayOut] (v4.center) \foreach \i in {4,3,2,1}{ -- (v\i)}	-- cycle;
	\draw[-,grayOut] (v1.center) \foreach \i in {12,11,10}{ -- (v\i)}	-- cycle;

	\draw[-] (v10) -- (v8);
	\draw[-] (v8) -- (v4) -- (v10);
	\draw[dashed,blue,very thick] (v4) -- (v9);
	
	\draw[-] (v8) node[label={-8*360/12:$k$},vertex2] {};
	\draw[-] (v9) node[label={-9*360/12:$k+1$},vertex2] {};
	\draw[-] (v10) node[label={-10*360/12:$k+2$},vertex2] {};
	
	\draw[-] (v1) node[label={-1*360/12:$i$},vertex2] {};
	\draw[-] (v4) node[label={-4*360/12:$j$},vertex2] {};

	\draw[-,polygonEdge] (v1.center) \foreach \i in {2,...,12}{-- (v\i.center)}	-- cycle;
\end{tikzpicture}
\caption{}
\end{subfigure}%
\begin{subfigure}[b]{0.25\textwidth}\centering
\begin{tikzpicture}
	\foreach \i in {1,...,12}{\node[vertex] (v\i) at (-\i*360/12:1.5) {};}
	
	\draw[-,grayOut] (v1.center) \foreach \i in {2,...,8}{-- (v\i.center)}	-- cycle;
	\draw[-,grayOut] (v1.center) \foreach \i in {12,11,10}{ -- (v\i)}	-- cycle;

	\draw[-] (v10) -- (v8);
	\draw[-] (v8) -- (v1) -- (v10);
	\draw[dashed,red2,very thick] (v1) -- (v9);

	\draw[-] (v8) node[label={-8*360/12:$k$},vertex2] {};
	\draw[-] (v9) node[label={-9*360/12:$k+1$},vertex2] {};
	\draw[-] (v10) node[label={-10*360/12:$k+2$},vertex2] {};
	
	\draw[-] (v1) node[label={-1*360/12:$i$},vertex2] {};
	\draw[-] (v4) node[label={-4*360/12:$j$},vertex2] {};

	\draw[-,polygonEdge] (v1.center) \foreach \i in {2,...,12}{-- (v\i.center)}	-- cycle;
\end{tikzpicture}
\caption{}
\end{subfigure}%
\begin{subfigure}[b]{0.25\textwidth}\centering
\begin{tikzpicture}
	\foreach \i in {1,...,12}{\node[vertex] (v\i) at (-\i*360/12:1.5) {};}
	
	\draw[-,grayOut] (v4.center) \foreach \i in {5,...,8}{-- (v\i.center)}	-- cycle;
	\draw[-,grayOut] (v4.center) \foreach \i in {4,3,2,1}{ -- (v\i)}	-- cycle;
	\draw[-,grayOut] (v1.center) \foreach \i in {12,11,10}{ -- (v\i)}	-- cycle;

	\draw[-] (v10) -- (v8);
	\draw[-] (v8) -- (v1) -- (v10);
	\draw[dashed,red2,very thick] (v1) -- (v9);
	
	\draw[-] (v8) node[label={-8*360/12:$k$},vertex2] {};
	\draw[-] (v9) node[label={-9*360/12:$k+1$},vertex2] {};
	\draw[-] (v10) node[label={-10*360/12:$k+2$},vertex2] {};
	
	\draw[-] (v1) node[label={-1*360/12:$i$},vertex2] {};
	\draw[-] (v4) node[label={-4*360/12:$j$},vertex2] {};

	\draw[-,polygonEdge] (v1.center) \foreach \i in {2,...,12}{-- (v\i.center)}	-- cycle;
\end{tikzpicture}
\caption{}
\end{subfigure}%
\caption{}
\label{fig:HedgeProof}
\end{figure}

Any other sub-triangulation of the gray region of (A) preserves $x_j$, so in particular one can choose the sub-triangulation with the pentagon $\st{k, k+1, k+2, i, j}$, shown in (B). Similarly, one can go from (C) to (D) and $x_i$ will still be accessible by the red chord flip. But now notice that this is exactly the situation from the $A_2$ case! There is an embedded pentagon with exactly the same triangulations that appeared above. Therefore $x_i - x_j$ factors as a product of cluster coordinates, i.e $x_i -x_j \in M_{A_n}$.
We can also show this algebraically, making use of a Pl\"ucker relation,
as displayed in eq.~\eqref{eq:clique_different_computation}.

\end{document}